\documentclass{article}
 \pdfoutput=1
\usepackage{graphicx}  
\usepackage{amsmath}   
\usepackage{amssymb}   
\usepackage[export]{adjustbox}
\usepackage{bm} 
\usepackage{dcolumn}
\usepackage{color}
\usepackage{mathrsfs}
\usepackage{amsfonts}
\usepackage{varioref}

\usepackage{physics}
\usepackage[vcentermath]{youngtab}
\usepackage{wrapfig}

\RequirePackage[colorlinks,citecolor=blue,urlcolor=magenta,linkcolor=blue]{hyperref}

\def\Tr{{\rm Tr}}

\labelformat{section}{Section #1} 
\labelformat{subsection}{Section #1} 
\labelformat{subsubsection}{Section #1}
\labelformat{subsubsubsection}{Section #1}
\labelformat{equation}{Eq.~(#1)} 
\labelformat{figure}{Fig.~#1} 
\labelformat{subfigure}{Fig.~\thefigure#1} 
\labelformat{table}{Tab.~#1} 
\labelformat{appendix}{Appendix #1}

\title{\bf Soft photon radiation and entanglement}
\author{\bf Anastasios Irakleous$^{1}$\footnote{irakleous.anastasios@ucy.ac.cy}, Theodore N Tomaras$^{2}$\footnote{tomaras@physics.uoc.gr}
\bf ~and~Nicolaos Toumbas$^{1}$\footnote{nick@ucy.ac.cy}\\
$^{1}$\small{Department of Physics, University of Cyprus, Nicosia 1678, Cyprus}\\
$^{2}$\small{ITCP and Department of Physics, University of Crete, 700 13 Heraklion, Greece}}

\begin{document}
  
\maketitle
\begin{abstract}
\noindent 
We study the entanglement between soft and hard particles produced in generic scattering processes in QED. The reduced density matrix for the hard particles, obtained via tracing over the entire spectrum of soft photons, is shown to have a large eigenvalue, which governs the behavior of the Renyi entropies and of the non-analytic part of the entanglement entropy at low orders in perturbation theory. The leading perturbative entanglement entropy is logarithmically IR divergent. The coefficient of the IR divergence exhibits certain universality properties, irrespectively of the dressing of the asymptotic charged particles and the detailed properties of the initial state. In a certain kinematical limit, the coefficient is proportional to the cusp anomalous dimension in QED. For Fock basis computations associated with two-electron scattering, we derive an exact expression for the large eigenvalue of the density matrix in terms of hard scattering amplitudes, which is valid at any finite order in perturbation theory. As a result, the IR logarithmic divergences appearing in the expressions for the Renyi and entanglement entropies persist at any finite order of the perturbative expansion. To all orders, however, the IR logarithmic divergences exponentiate, rendering the large eigenvalue of the density matrix IR finite. The all-orders Renyi entropies (per unit time, per particle flux), which are shown to be proportional to the total inclusive cross-section in the initial state, are also free of IR divergences. The entanglement entropy, on the other hand, retains non-analytic, logarithmic behavior with respect to the size of the box (which provides the IR cutoff) even to all orders in perturbation theory.   
\end{abstract}
\vskip .5cm 
\noindent
\noindent
{\bf keywords : Entanglement entropy, Renyi entropies, soft theorems, IR divergences} 
\bigskip
\setcounter{tocdepth}{2}

\section{Introduction}\label{s1}

The emission of soft radiation is ubiquitous in scattering processes in gauge theories and gravity. The hard asymptotic particles are accompanied by an infinite number of soft photons and gravitons. Much of the structure of the IR dynamics is dictated by symmetry, independently of the details of the UV completion of the theory \cite{StromingerLectures,StromingerIRrevisited,StromingerBMS,HMPS,LPS,KMS,SZ,Campiglia,Sever,Weinberg}. It is important to provide measures of the entanglement between the hard and soft degrees of freedom \cite{Carney1,Carney2,Carney3,Carney4,Gomez1,Gomez2,TT}, and to uncover implications for the structure of the S-matrix (in both gauge theories and gravity) \cite{Arkani,Pate,Himwich,Ding,Banerjee1}, black hole physics \cite{HPS,StromingerBHinfo,Porrati,HHPS,Pasterski} and holography \cite{Banks1,Banks2,Guevara,Banerjee2}.    

In \cite{TT}  we computed the leading perturbative entanglement entropy between the soft and the hard particles produced in two-electron scattering processes in QED. The initial and final states were dressed with clouds of soft photons, in accordance with the Faddeev-Kulish construction \cite{FK,Chung,Kibble1,Kibble2}. We found that tracing over the entire spectrum of soft photons, including those in the clouds, leads to (logarithmic) IR divergences in the perturbative expansion of the Renyi and the entanglement entropies. The coefficient of the logarithmic divergence was found to be independent of the dressing and to contain physical information. In a certain kinematical limit, it is related to the cusp anomalous dimension in QED \cite{Korchemsky}.

A question that arises concerns the order of limits. We place the system in a large box of size $L$, imposing, therefore, an infrared cutoff $\lambda$ of order $1/L$. For the perturbative calculations, we first expand to a given order in perturbation theory, keeping $\lambda$ finite, and take the continuum, $\lambda \to 0$ limit at the end. The logarithmically divergent quantities in this limit are the entanglement and Renyi entropies per unit time per particle flux. 

Thus, because of the IR divergences, the perturbative expansion of the entanglement and Renyi entropies breaks down in the strict $\lambda \to 0$ limit. Rather, its validity is guaranteed at small, but fixed $\lambda$, taking the electron charge $e$ to be sufficiently small. Equivalently, we take the size of the box to be large but finite.\footnote{It was suggested in \cite{Gomez1,Gomez2,TT} to distinguish between the soft cloud photons, ``dressing'' the asymptotic charged particles, and additional radiative photons, comprising the soft part of the emitted radiation. The extra soft radiated photons have energies greater than the energy characterizing the soft cloud photons. Indeed, the dressed amplitudes describing the emission of photons with energy less than or equal to the energy of the photons present in the clouds are suppressed. Restricting the partial trace over the additional soft radiated photons leads to a well-behaved perturbative expansion for the Renyi and entanglement entropies. In particular, these are finite, free of IR divergences, order by order in perturbation theory \cite{TT}. We will not study this second type of partial tracing in this work.}          

These results suggest that the continuum, $\lambda \to 0$ limit does not commute with the perturbative expansion. An analogy to keep in mind concerns the perturbative expansion of the conventional Fock basis scattering amplitudes. At any finite order in perturbation theory, these amplitudes are nonvanishing and exhibit IR divergences due to virtual soft photons. To all orders, however, these IR divergences exponentiate leading to the vanishing of these amplitudes \cite{Yennie,Weinberg}. This result can also be understood as a consequence of symmetries associated with large gauge transformations \cite{StromingerLectures,StromingerIRrevisited}.

The structure of the reduced density matrix is indeed very suggestive. At any finite order in perturbation theory, the matrix elements that are off-diagonal  with respect to momentum indices, are non-zero, containing IR logarithmic divergences in $\lambda$ (for both the dressed and undressed cases). To all orders in the electron charge however, the logarithmic IR divergences exponentiate, leading to the vanishing of these off-diagonal elements in the continuum limit \cite{Carney2,Carney3,Carney4}. The diagonal elements are given in terms of inclusive Bloch-Nordsieck rates associated with dressed box states. These are free of any IR divergences in $\lambda$, order by order in perturbation theory \cite{Bloch,Yennie,Weinberg}. However they scale inversely proportional with powers of the volume of the box. They also tend to zero in the continuum limit, albeit less fast than the off-diagonal elements in the momentum. 

So, to all orders in perturbation theory, the reduced density matrix assumes an almost diagonal form in the continuum limit -- there are nonvanishing off-diagonal elements with respect to particle polarization indices. The number of nonvanishing eigenvalues grows large, revealing strong entanglement between the soft and hard degrees of freedom and decoherence \cite{Carney1,Carney2,Carney3,Carney4,Gomez1,Gomez2}. See also \cite{Petruccione, Calucci} for earlier work. 

As we will see, there is an exponentiation pattern for the IR divergences appearing in the perturbative expansion of the Renyi entropies. As a result, {\it to all orders in the electron charge,} the Renyi entropies are free of logarithmic IR divergences, rendering the continuum limit well defined. The entanglement entropy, though, has non-trivial volume dependence. It is non-analytic in terms of dimensionless parameters given by products of the size of the box $L$ and various energy scales characterizing the scattering process. 

Below we outline the plan and summarize the main results of this paper. 

We first generalize the perturbative computation of \cite{TT} to generic scattering processes, and study the structure of the leading entanglement entropy. We consider an initial Faddeev-Kulish state with arbitrary (but relatively small) numbers of electrons and positrons, and trace over the entire spectrum of soft photons in the final state. The reduced density matrix has one large eigenvalue, governing the behavior of the Renyi entropies for integer $m\ge 2$ and the entanglement entropy at leading order. The coefficient of the IR logarithmic divergence of the leading entanglement entropy exhibits certain universality properties. In particular, it is independent of the dressing of the incoming charged particles with soft photons. The dominant contributions arise from amplitudes associated with two-particle interactions, at small scattering angle or scattering angle close to $\pi$. In the relativistic, high energy limit, these dominant contributions are proportional to the cusp anomalous dimension in QED, extending the result of \cite{TT} to more general scattering processes. We also find that the entanglement entropy per unit time per particle number density scales proportionally with the number of charged particles in the initial state. These results are described in great detail in \ref{s2}. 

We proceed in \ref{s3} to compute the next to leading order corrections to the Renyi and entanglement entropies. Since the leading order analysis reveals that the singular part of the entanglement entropy is independent of the dressing function, we focus in a Fock basis computation concerning electron-electron scattering for simplicity. At next to leading order, two types of IR divergent terms appear in the expressions for Renyi entropies. The first is proportional to the sum of the squares of the amplitudes for the emission of two soft photons, and it is of order $(\ln\lambda)^2$.  The second is proportional to the sum of the squares of the amplitudes for the emission of one soft and one hard photon, and it is of order $\ln\lambda$. A similar term to the latter one appears at leading order, but the coefficient is controlled by the sum of the squares of the amplitudes for single soft photon emission \cite{TT}. We comment on the pattern of the exponentiation of these logarithmic IR divergences. In addition, we obtain the next to leading order corrections to the coefficient of the non-analytic part of the entanglement entropy.  

The analysis of the two-electron scattering case {\it to all orders in perturbation theory} is carried out in \ref{s4}. 
We derive an exact expression for the large eigenvalue of the density matrix in terms of a series of hard scattering amplitudes. At any finite order in perturbation theory, this eigenvalue is plagued with logarithmic IR divergences due to virtual soft photons. To all orders, however, the IR divergences exponentiate leading to a finite, nonvanishing result. The shift of the large eigenvalue from unity becomes arbitrarily small in the continuum, large volume limit, determining completely the behavior of the Renyi entropies to all orders. The contributions from the small nonzero eigenvalues (to the Renyi entropies) remain subleading, despite the fact that their number grows with the volume of the box. Thus the Renyi entropies to all orders are free from any IR divergences in the continuum limit. The Renyi entropies per unit time per particle flux are proportional to the total, inclusive cross-section in the initial sate. We comment on the dependence of the total cross-section on the reference energy scale separating the soft from the hard parts of the Hilbert space, as well as other characteristic energy scales associated with the scattering process. The entanglement entropy, on the other hand, retains the non-analytic, logarithmic behavior with respect to the size of the box $L$, even to all orders in the continuum limit.      

Throughout, we follow the notations and conventions of \cite{TT} -- see also \ref{a1}. 
The following leading soft theorems \cite{Weinberg,Yennie,Bloch,Low1,GellMann,Low2,BK,Duca} concerning soft photon emission are applied
\begin{equation}
\lim_{|\vec{q}|\to 0}~(2\omega_{\vec{q}})^{1/2}~\langle\beta|a_r(\vec{q})~S~|\alpha\rangle=\left(\sum_{i\in\beta}~ \frac{e_i\,p_i \cdot \epsilon_r^*(\vec{q})}{p_i \cdot q}~-~\sum_{i\in \alpha}~ \frac{e_i\, p_i \cdot \epsilon_r^*(\vec{q})}{p_i \cdot q}\right)~\langle\beta|~S~|\alpha\rangle
\end{equation}
and 
\begin{equation}
\lim_{|\vec{k}|\to 0}~(2\omega_{\vec{k}})^{1/2}~\langle\beta|~S~a^{\dagger}_r(\vec{k})~|\alpha\rangle=-\left(\sum_{i\in\beta}~ \frac{e_i\,p_i \cdot \epsilon_r(\vec{k})}{p_i \cdot k}~-~\sum_{i\in \alpha}~ \frac{e_i\, p_i \cdot \epsilon_r(\vec{k})}{p_i \cdot k}\right)~\langle\beta|~S~|\alpha\rangle
\end{equation}

A multielectron/positron dressed state $\alpha=\{e_i, \vec{p}_i, s_i\}$ is given as a product of a conventional Fock basis state $\ket{\alpha}$ and a coherent state $|f_\alpha\rangle$ describing the photons in the cloud 
\begin{equation}
 |\alpha\rangle_{d}= |\alpha\rangle\times|f_\alpha\rangle
\end{equation} 
where
\begin{equation}
|f_\alpha\rangle=e^{\int_\lambda^{E_d}\frac{d^3\vec{k}}{(2\pi)^3}~\frac{1}{(2\omega_{\vec{k}})^{1/2}}~\left(f_{\alpha}(\vec{k},\,\vec{p})\cdot a^{\dagger}(\vec{k}) - h.c.\right)}|0\rangle\label{coherent}
\end{equation}
The dressing function is given by \cite{FK,Chung}
\begin{equation}
f_\alpha^{\mu}(\vec{k})=\sum_{i \in \alpha}~e_i~\left(\frac{p_i^{\mu}}{p_ik} - c^{\mu}\right)~e^{-ip_ik\, t_0/p_i^0} 
\label{dressing}
\end{equation}
In particular, the dressing function $f_{\alpha}^{\mu}(\vec{k},\,\vec{p})$ is singular as the photon momentum $\vec{k}$ vanishes.  Also, $t_0$ is a time reference scale and $c^{\mu}$ is a null vector, $c^2=0$, satisfying $ck=1$.  Only soft photons, with energies below the infrared reference scale $E_d$, are present in the cloud. Taking $E_d < 1/t_0$, we may approximate the phase $e^{-ipk\, t_0/p^0}$ in \ref{dressing} with unity -- see \cite{Gomez2} for comprehensive discussions. S-matrix elements between asymptotic dressed states are free of IR logarithmic divergences, order by order in perturbation theory \cite{FK,Chung,Kibble1,Kibble2,Choi1,Choi2,Sotaro}. The dressing energy scale can be taken to be sufficiently small so as the leading soft theorems can be readily applied in various computations of amplidutes. 
 
In \cite{TT} we computed the overlap between coherent photon states associated with generic charged states $\alpha=\{e_i,\, \vec{p}_i,\,s_i\}$ and $\beta=\{e_i^{\prime}, \,\vec{p}_i^{\,\prime},\,s_i^{\,\prime}\}$. See also \cite{Gomez2}. Calling the $\beta$ particles outgoing and the $\alpha$ particles incoming, we define $\eta_i$ to be $+1$ for all outgoing particles and $-1$ for all incoming particles. Then when the net charges of the incoming and outgoing states are equal, $Q_\alpha=Q_\beta$, and to all orders in the electron charge $e$, we find 
\begin{equation}
\langle f_\beta|f_\alpha\rangle = \left(\frac{\lambda}{E_d}\right)^{{\cal{B}}_{\beta\alpha}}\label{braketf}
\end{equation}
where
\begin{equation}
{\cal{B}}_{\beta\alpha}=-\frac{1}{16\pi^2}~\sum_{ij}~\eta_i\,\eta_j\,e_i\,e_j~v_{ij}^{-1}~ \ln \left(\frac{1+ v_{ij}}{1-v_{ij}}\right)\label{B}
\end{equation}
and 
\begin{equation}
v_{ij}=\left[1-\frac{m_i^2\,m_j^2}{(p_i\cdot p_j)^2}\right]^{1/2}
\end{equation}
is the magnitude of the relative velocity of particle $j$ with respect to $i$.
The overlap admits the following expansion
\begin{equation}
\bra{f_\beta}f_\alpha\rangle=e^{{\cal{B}}_{\beta\alpha}\ln(\lambda/E_d)}=1+{\cal{B}}_{\beta\alpha}\ln(\lambda/E_d)+\dots \label{bracketf2}
\end{equation} 
Notice that when the momenta of the states $\beta$ and $\alpha$ differ, ${\cal{B}}_{\beta\alpha}$ is nonzero and positive \cite{Weinberg}. Therefore, to all orders in the electron charge, the overlap vanishes in the $\lambda \to 0$ limit.

As we already remarked, the infrared cutoff $\lambda$ is taken naturally to be the inverse of the size of the box: $\lambda = 2\pi/L$. The particle momenta take discrete values and the corresponding single particle states are unit normalized. 

Using a reference energy scale $E$, which we take to be smaller than the mass of the electron, we separate the asymptotic Hilbert spaces into soft and hard factors 
\begin{equation}
\mathcal{H}=\mathcal{H}_H\times\mathcal{H}_S  
\end{equation}
where $ \mathcal {H}_H $ includes particle states with energy greater than $E$ and $ \mathcal {H}_S $ includes photon states with total energy less than $E$. We take the energy scale $E_d $, characterizing the cloud photons to be arbitrarily small, so that we can apply soft theorems to simplify various expressions:  $\lambda <E_d <E$.  We take the scale $\Lambda$ characterizing soft virtual photons to be of the order of $E_d$ (or $E$ for Fock basis calculations). The continuum limit, $ \lambda \rightarrow 0$, is taken at the end of the computation, keeping the ratios $E_d/E$, $E_d/\Lambda$ constant. For the Fock basis computations, we set the dressing function \ref{dressing} to be zero. The relevant energy scales in this latter case are $\lambda$, $\Lambda$ and $E$.
Further work on entanglement after scattering includes \cite{Bala,Seki,Grignani,Asorey,Rai}.

\section{Scattering and entanglement}\label{s2}
\setcounter{equation}{0}
In this section we extend the perturbative analysis of \cite{TT} to generic scattering setups. We will work at leading order in perturbation theory. Higher order corrections are discussed in the following sections. 

Consider an incoming Faddeev-Kulish state $\alpha$ with $m$ electrons and $n$ positrons. The corresponding bare state $\ket{\alpha}$ is taken to be a momentum eigenstate. Then the incoming state factorizes in $\mathcal{H}_H\times\mathcal{H}_S$ as follows
\begin{equation}  
|\Psi\rangle_{in}=|\alpha\rangle_{d}= |\alpha\rangle_{H}\times|f_\alpha\rangle_{S} 
\end{equation} 
Initially, there is no entanglement between the soft and hard degrees of freedom since $\ket{\alpha}_d$ is a product state. Entanglement arises as a result of scattering. In order to ensure the applicability of the perturbative analysis, we keep the total number of charged particles in the initial state to be relatively small. 

The outgoing state is given by
\begin{equation}  \label{Smatrix}
|\Psi\rangle_{out}=S|\Psi\rangle_{in}= (1+iT)|\alpha\rangle_{d}
\end{equation}
where we denote by $iT$ the non-trivial part of the S-matrix.
Since the S-matrix is unitary, the following relation holds
\begin{equation}  \label{unitarity}
i(T-T^{\dagger})+T^{\dagger}T=0
\end{equation}
Inserting a complete basis of dressed states, the outgoing state can be written as 
\begin{equation} \label{Psiout}
|\Psi\rangle_{out}=|\alpha\rangle_{d}+ \widetilde{T}_{\beta\alpha}|\beta\rangle_{d}+\widetilde{T}_{\beta\gamma,\alpha}|\beta\gamma\rangle_{d}+\dots
\end{equation}
where $\widetilde{T}_{\beta\alpha}=_d\langle\beta|iT|\alpha\rangle_{d}$ and $\widetilde{T}_{\beta\gamma,\alpha}=_d\langle\beta\gamma |iT|\alpha\rangle_{d}$ are S-matrix elements between dressed states. Summation over the final state indices $\beta$ and $\gamma$ is implied. The indices stand for both the momenta and the polarizations of the particles.

The set $\{|\beta\rangle\}$ includes all allowable final states with no photons. In addition, it includes final states with the smallest number of photons that can be produced as a result of a given number of electron/positron annihilations. These photons are hard.\footnote{The momenta of the incoming particles are taken to be different and generic. In order to get one of the emitted photons to be soft, we must consider very special cases in which the electron ``chases'' the positron. We do not consider such special cases in this work.} For example, if $l$ annihilations take place, the corresponding state contains $m-l$ electrons, $n-l$ positrons and $2l$ hard photons.  Notice that $0\le l\le \rm{min}(m,n)$.
The state $|\beta\gamma\rangle$ contains an additional photon, which can be either hard or soft, emitted from a charged particle.

The undressed amplitudes $T_{\beta\alpha}$ and $T_{\beta\gamma,\alpha}$\footnote{We would like to stress that in our notation $T_{x,y}\, (\widetilde T_{x,y})$ are the matrix elements of the operator $i T$ between undressed and dressed states, respectively.} are given in terms of the conventional connected and disconnected Feynman diagrams. The leading contributions to $T_{\beta\alpha}$ arise from disconnected diagrams (or connected in some special cases) of order $e^{2}$ if no photons are present in $|\beta\rangle$, and of order $e^{2l}$ if $l$ pairs of photons are present. Accordingly, the leading contribution to ${T}_{\beta\gamma,\alpha}$ is of order $e^{3}$ or $e^{2l+1}$. We denote the order of  $T_{\beta\alpha}$ with $e^{\cal{N}}$ (where ${\cal{N}}$ depends on the details of $\beta$). 

We will also use the following relations between dressed and undressed amplitudes \cite{Chung,Gomez2,TT}
\begin{equation}\label{du1}
{\widetilde T}_{\beta\alpha}=T_{\beta\alpha}/\bra{f_\beta}\ket{f_\alpha}
\end{equation}
and
$$
{\widetilde T}_{\beta\gamma,\,\alpha}=T_{\beta\gamma,\,\alpha}/\bra{f_\beta}\ket{f_\alpha}~~~~{\rm{if}}~~~\omega_\gamma>E_d
$$
\begin{equation}\label{du2}
{\widetilde T}_{\beta\gamma,\,\alpha} ~=~\frac{1}{(2V\omega_\gamma)^{1/2}}~F_{\beta\alpha}(\vec{q}_\gamma,\,\epsilon_r(\vec{q}_\gamma))~~~~{\rm{if}}~~~\omega_\gamma<E_d
\end{equation}
As shown in \cite{Choi2}, the function $F_{\beta\alpha}(\vec{q}_\gamma,\,\epsilon_r(\vec{q}_\gamma))$ is smooth and nonsingular in the limits $\lambda,\,\omega_\gamma=|\vec{q}_\gamma| \to 0$ (and of order the dressing scale $E_d$). The volume factor appears due to the relative normalization between box and continuum states (see \ref{cl}). 

The ellipses in \ref{Psiout} stand for higher order contributions, which do not contribute to the leading Renyi and entanglement entropies. 
The outgoing density matrix (which is pure) is $|\Psi\rangle_{out}\langle\Psi|_{out}$.

To obtain the hard density matrix $\rho_H$, we trace over the entire spectrum of soft degrees of freedom. These include cloud photons with energies less than the dressing scale $E_d$, and additional radiative photons with energy less than 
$E$.\footnote{The dressed amplitudes for photon emission are suppressed when the photon energy is below $E_d$ \cite{Choi2}.} Therefore, the reduced density matrix is given by
\begin{equation}  \label{dHard}
\rho_{H}=\Tr_{H_S}\left(|\Psi\rangle_{out}\langle\Psi|_{out}\right)
\end{equation}
Using \ref{Psiout} and implementing the partial trace, we can expand $\rho_H$ in terms of ket-bra operators associated with conventional, Fock-basis hard states. The terms relevant for the leading order computation of the Renyi and the entanglement entropies are derived explicitly in \ref{a2}. A useful list of formulae to calculate the various partial traces is included in \ref{a1}. 

Let us stress that for our perturbative calculations, we expand to a given order in perturbation theory, keeping the volume of the box finite, and take the continuum $\lambda \to 0$ limit at the end. The matrix elements of $\rho_H$ are given in terms of dressed amplitudes, which are free of IR divergences in $\lambda$ (at least perturbatively), as well as overlaps of coherent photon states describing the clouds -- see \ref{reduced}. The diagonal elements are proportional to inclusive Bloch - Nordsieck rates associated with dressed box states. They are free of any IR divergences in $\lambda$, order by order in perturbation theory \cite{TT}. The off diagonal elements are proportional to the overlaps $\langle f_{\beta^{\prime}}|f_{\beta}\rangle$, which at any finite order in perturbation theory, induce logarithmic divergences in $\lambda$ -- see \ref{bracketf2}. Finally, the unitarity relation \ref{unitarity} ensures that $\Tr\rho_{H}=\Tr\rho=1$. 

\subsection{The large eigenvalue of $\rho_H$ and the leading order Renyi entropies}
In \ref{a2}, we compute the Renyi entropies 
\begin{equation}  \label{SRenyi}
S_{m}=\frac{1}{1-m}\ln\Tr(\rho_H)^{m}
\end{equation}
for integer $m\geq2$ perturbatively, extending the analysis of \cite{TT} to the generic case. The leading order Renyi entropies are of order $e^6$, irrespectively of the number of charged particles in the initial state. This is because disconnected diagrams contribute to leading order. Either a pair of electrons or positrons scatter and the rest propagate intact, or an electron/positron pair gets annihilated to two photons (and the rest of the charged particles propagate intact). 

Indeed to leading order, we find for $m\geq 2$
\begin{equation}   
\Tr(\rho_H)^m=1-m\Delta
\end{equation}
where
\begin{equation}   \label{Delta}
\Delta=\sum_{\beta}\sum_{\omega_\gamma<E} T_{\beta\gamma,\alpha}T^{*}_{\beta\gamma,\alpha}
\end{equation}
is an order $e^6$ quantity, depending on the amplitude for single real photon emission in the energy range $\lambda<\omega_\gamma<E$. Therefore, the Renyi entropies for integer $m\ge 2$ take the form
\begin{equation}   
S_{m}=-\frac{1}{m-1}\ln[1-m\Delta] =\frac{m}{m-1}\Delta
\end{equation}

These results are in accordance with the fact that $\rho_H$ has one large eigenvalue, which is equal to $1-\Delta$ to order $e^6$. All other nonvanishing eigenvalues are of order $e^6$ (or higher). Their sum must be equal to $\Delta$. To prove this, let us set 
\begin{equation}\label{Phi}
\ket{\Phi}= |\alpha\rangle_{H}~+~\sum_{\beta}C_{\beta} |\beta\rangle_{H}~+~\sum_{\beta}\sum_{\omega_\gamma>E}C_{\beta\gamma} |\beta\gamma\rangle_{H}~+~\dots 
\end{equation}
as dictated by the first two lines of  \ref{reduced}, and write the reduced density matrix in the form
\begin{equation}
\rho_H=\ket{\Phi}\bra{\Phi}+G
\end{equation} 
where $G$ is an order $e^6$ matrix.\footnote{The ellipses in \ref{Phi} stand for contributions from states containing more than one photons, which do not affect the matrix elements of the perturbation $G$ and the eigenvalue analysis to leading order.} The coefficients $C_\beta$ and $C_{\beta\gamma}$ are given in terms of dressed amplitudes and overlaps of coherent cloud photon states in \ref{Cbeta} and \ref{Cbetagamma}, respectively. Specifically we obtain
\begin{equation}
G=\sum_{\beta,\beta'}~\bigg[\left(\bra{f_{\beta'}}f_\beta\rangle-\bra{f_\alpha}f_\beta\rangle\bra{f_\alpha}f_{\beta'}{\rangle}^*\right)\widetilde{T}_{\beta\alpha}\widetilde{T}^*_{\beta'\alpha}+\bra{f_{\beta'}}\ket{f_\beta}\sum_{\omega_{\gamma<E}}\widetilde{T}_{\beta\gamma,\alpha}\widetilde{T}^*_{\beta'\gamma,\alpha}\bigg]\ket{\beta}\bra{\beta'}~+~\dots
\end{equation}
where the ellipses stand for higher order terms. Using this expression (as well as energy conservation), we find that $G$ annihilates $\ket{\Phi}$ to order $e^6$.\footnote{Up to negligible terms, of order $E_d^2$ (at most), which we drop.}  Thus  $\ket{\Phi}$ is an eigenstate of $\rho_H$ with eigenvalue
\begin{equation}
\lambda_{\Phi}=1-\Delta~+~{\cal{O}}(e^8)
\end{equation}
The last equality is shown explicitly in \ref{a2}. The other nonvanishing eigenvalues coincide with the nonvanishing eigenvalues of the matrix $G$.\footnote{The corresponding eigenstates must be orthogonal to $\ket{\Phi}$.} Each is at least of order $e^6$. Their sum is equal to $\Delta$, as can be verified by computing explicitly the trace of $G$ to order $e^6$, \ref{G1} and \ref{G2}.

\subsection{Entanglement entropy to leading order}

The entanglement entropy is given by the expression
\begin{equation}  \label{Sent}
S_{ent}=-\Tr\rho_{H}\ln\rho_H=-\sum_i \lambda_i \ln{\lambda_i}
\end{equation}
where the sum runs over the eigenvalues of the reduced density matrix. As we have seen, to leading perturbative order ($e^6$), there is a ``large'' eigenvalue, $\lambda_\Phi = 1- \Delta$, and the rest non-zero eigenvalues add up to $\Delta$. We set $\lambda_i = e^6 a_i$ when $i\neq \Phi$, where $a_i$ are order one quantities, to obtain
\begin{equation}
S_{ent}=-\Delta \ln{e^6}+\Delta-e^6\sum_{i \neq \Phi}a_i\ln{a_i}=-\Delta \ln{e^6}+{\cal{O}}(e^6)
\end{equation}
The leading entanglement entropy is of order $e^6\ln{e^6}$. 

The coefficient of the non-analytic part is obtained to be equal to $\Delta$.
We will consider the dominant singular part in the limit $\lambda\rightarrow0$. We get
\begin{equation}   
\Delta_{sing}=\sum_{\beta}\sum_{\omega_\gamma<E_d}T_{\beta\gamma,\alpha}T^{*}_{\beta\gamma,\alpha}
\end{equation} 

Using soft theorems, we obtain for the singular part of the leading entanglement entropy
\begin{equation}   \label{Sentsing}
S_{ent,sing}=-\ln{e^6}\,\sum_{\beta}(T_{\beta\alpha}T^{*}_{\beta\alpha})\bigg(\sum_{\omega_\gamma<E_d}\frac{1}{2V\omega_\gamma}\sum_{ss^{'}\in\{\alpha,\beta\}}e_{s}e_{s^{'}}\eta_{s}\eta_{s^{'}}\frac{p_{s} p_{s^{'}}}{(p_{s}q_{\gamma})(p_{s^{'}}q_{\gamma})} \bigg)
\end{equation} 
where the bare amplitude $T_{\beta\alpha}$ is calculated at tree level. The sum over $\beta$ is now restricted to final states for which the leading non-trivial (disconnected) diagrams are of order $e^2$. {\it This formula is universal, applicable for generic initial states. It is also independent of the dressing.}

\subsection{Continuum limit}\label{cl}

To take the continuum limit, we need the relative normalization between continuum states and box states
\begin{equation}   \label{bcnorm}
|\vec{p}\rangle_{Box}=\frac{1}{(2E_{\vec{p}}V)^{1/2}}|\vec{p}\rangle
\end{equation} 
The relevant final states $\{\ket{\beta}\}$ comprise configurations of  $m$ electrons and $n$ positrons, arising from the scattering of any two charged particles in the initial state, as well as configations of $m-1$ electrons, $n-1$ positrons and $2$ photons, resulting from the annihilation of an electron/positron pair. In either case the number of particles in the final state is $N=m+n$. We denote the corresponding invariant amplitudes by $i{\cal{M}}(\alpha \to me^-+ne^+)$ and $i{\cal{M}}(\alpha\to (m-1)e^-+(n-1)e^++2\gamma)$, respectively. We shall consider generic cases where the momenta of the initial particles differ from each other. For simplicity we average over the polarizations of the initial particles.

The singular part of the leading entanglement entropy takes the following form in the continuum limit
$$
S_{ent,sing}=-\frac{1}{V^N}\frac{\ln{e^6}}{m!\,n!}\Bigg\{\bigg(\prod_{f=1}^N\int\frac{d^3\vec{p}_f}{(2\pi)^32E_f}\bigg)\bigg(\prod_{i=1}^N\frac{1}{2E_i}\bigg) 
$$
$$
\times\left(|i\mathcal{M}_{\alpha}^{(me^-,ne^+)}|^2+\frac{mn}{2}|i\mathcal{M}_{\alpha}^{((m-1)e^-,(n-1)e^+,2\gamma)}|^2\right)\Big[(2\pi)^4\delta^4\Big(\sum_fp_f-\sum_ip_i\Big)\Big]^2
$$
\begin{equation}
\times\int_{\lambda}^{E_d}\frac{d^3\vec{q}}{(2\pi)^32\omega_{\vec{q}}}\sum_{s,r\in\{\alpha,\beta\}}\eta_s\eta_re_se_r\frac{p_sp_r}{(p_sq)(p_rq)}\Bigg\}
\end{equation}
The integral over the soft photon momentum can be carried out using the relation
\begin{equation} 
\int_{\lambda}^{E_d}\frac{d^3\vec{q}}{(2\pi)^32\omega_{\vec{q}}}\sum_{s,r\in\{\alpha,\beta\}}\eta_s\eta_re_se_r\frac{p_sp_r}{(p_sq)(p_rq)}=\ln\left(\frac{E_d}{\lambda}\right)\, 2\mathcal{B}_{\beta\alpha}
\end{equation}
yielding a logarithmically divergent factor in the IR cutoff $\lambda$.  The kinematical factor $\mathcal{B}_{\beta\alpha}$ is positive, given in terms of the momenta of the initial and final charged particles by \ref{B}.
Therefore, the leading entanglement entropy takes the form
$$
S_{ent,sing}=-\frac{2}{V^N}\frac{\ln{e^6}}{m!\,n!} \ln\left(\frac{E_d}{\lambda}\right) \Bigg\{\bigg(\prod_{f=1}^N\int\frac{d^3\vec{p}_f}{(2\pi)^32E_f}\bigg)\bigg(\prod_{i=1}^N\frac{1}{2E_i}\bigg) 
$$
\begin{equation}\label{SENTCONT}
\times\left(|i\mathcal{M}_{\alpha}^{(me^-,ne^+)}|^2+\frac{mn}{2}|i\mathcal{M}_{\alpha}^{((m-1)e^-,(n-1)e^+,2\gamma)}|^2\right)\mathcal{B}_{\beta\alpha}
\Big[(2\pi)^4\delta^4\Big(\sum_fp_f-\sum_ip_i\Big)\Big]^2\Bigg\}
\end{equation}
This expression generalizes the two-electron scattering case studied in \cite{TT}. Notice also that it is Lorentz invariant. We would like to understand the scaling with the number of charged particles $N$ in the initial state. 

\begin{figure}\label{3particle}
\begin{center}
\includegraphics [scale=1.00]{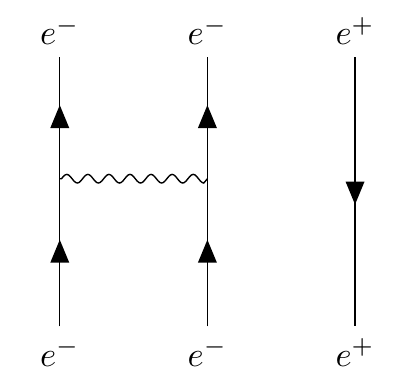}
\,\,\,\,\,\,\,\,\,\,\,\,\,\,\,\,\,\,\,\,\,\,\,\,\,\,\,\,\,\,\,\,\,
\includegraphics [scale=0.22]{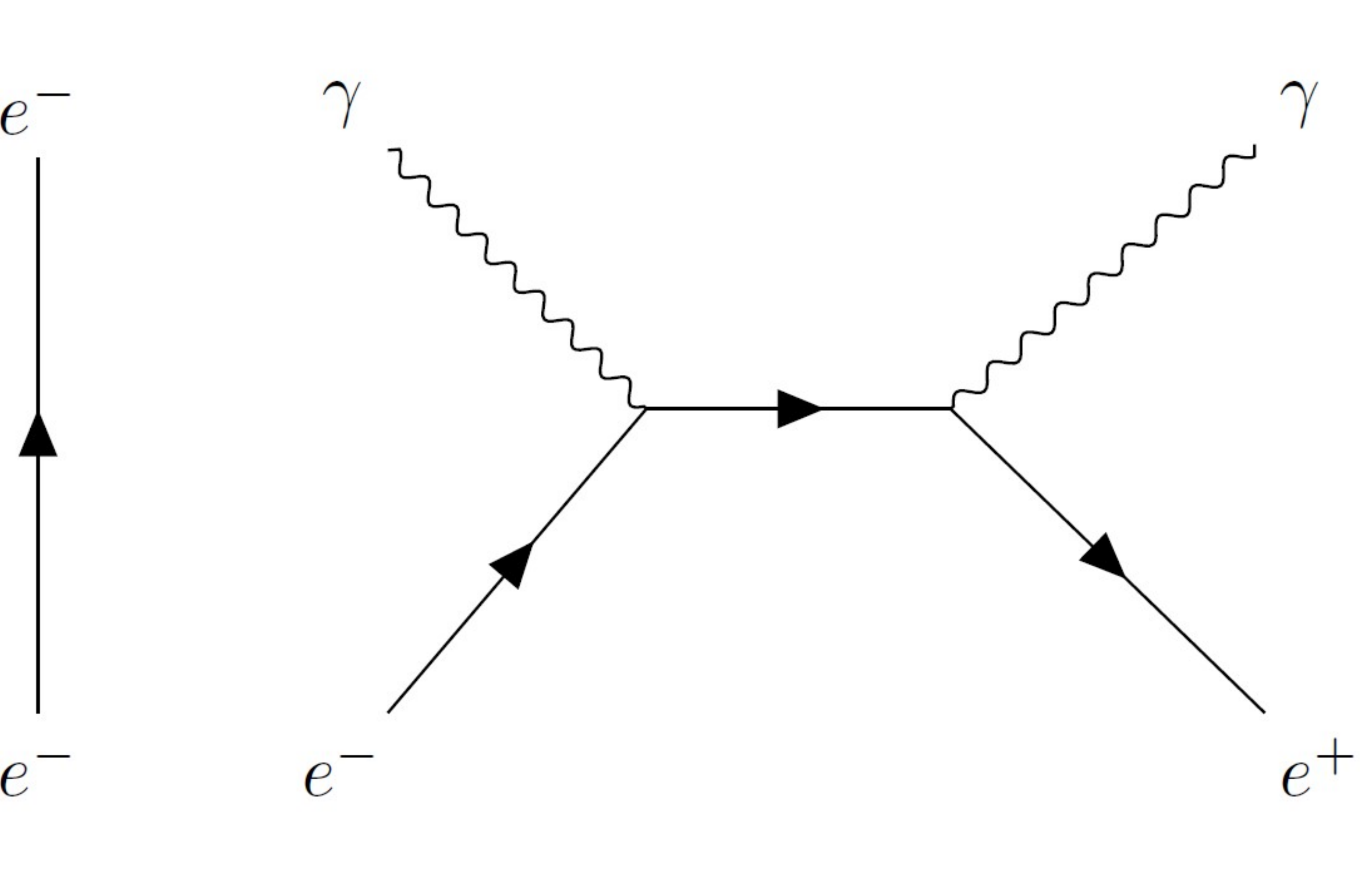}
\caption{\it \footnotesize Non-trivial disconnected Feynman diagrams associated with a two electron / positron initial state contributing to the leading entanglement entropy.}
\end{center}
\end{figure}

In the generic case, non-trivial disconnected diagrams in which only two particles interact with each other contribute at leading order. In \ref{a3} we explain in detail how to obtain the dominant leading contributions to the squares of the amplitudes $i{\cal{M}}(\alpha \to me^-+ne^+)$ and $i{\cal{M}}(\alpha \to (m-1)e^-+(n-1)e^++2\gamma)$ in the continuum, large volume limit. We also implement the integrations over the momenta of the final particles in \ref{SENTCONT}. 

Disconnected Feynman diagrams for a three particle initial state that contribute to the entanglement entropy at leading order are shown in figure 1. Detailed diagrams and explicit expressions are included in \ref{a3}.

We end up with the following expression for the leading entanglement entropy 
$$
S_{ent,sing}=-\frac{T\,\ln{e^6}}{32\pi\,V}\ln\left(\frac{E_d}{\lambda}\right) 
$$
$$
\times\, \int_{0}^\pi d\theta \sin{\theta}\, \Bigg\{\,\sum_{i,j=1}^m  \frac{v_{ij}}{2E_{ij}^2} |i\mathcal{M}_1(E_{ij}, \theta)|^2\mathcal{B}_1({i'j';ij}) + \sum_{i,j=1}^n  \frac{v_{ij}}{2E_{ij}^2} |i\mathcal{M}_2(E_{ij}, \theta)|^2\mathcal{B}_2({i'j';ij}) 
$$
\begin{equation}\label{entmaster}
+ 2\sum_{i=1}^m\sum_{j=1}^n  \frac{v_{ij}}{2E_{ij}^2} |i\mathcal{M}_3(E_{ij}, \theta)|^2\mathcal{B}_3({i'j';ij})+ \sum_{i=1}^m\sum_{j=1}^n  \frac{v_{ij}}{2E_{ij}^2} |i\mathcal{M}_4(E_{ij}, \theta)|^2\mathcal{B}_4({i'j';ij})\Bigg\} 
\end{equation}
Here, $T$ is the time scale of the experiment (arising from an overall energy conserving $\delta$-function). The sums run over all possible pairs of interacting incoming particles. We may always choose to work in the center of mass frame of the two interacting particles, where $\vec{p}_i=-\vec{p}_j$. The scattering angle is denoted by $\theta$ and the center of mass energy by $E_{ij}=2E_i$. The magnitude of the relative velocity is given by $v_{ij}=4|\vec{p}_i|/E_{ij}$.  We denote the scattering amplitudes for the four relevant processes $e^-e^-\to e^-e^-$, $e^+e^+\to e^+e^+$, $e^-e^=\to e^-e^+$ and $e^+e^-\to \gamma\gamma$ by $\mathcal{M}_1, \mathcal{M}_2,\mathcal{M}_3,\mathcal{M}_4$, respectively. Finally, $\mathcal B_1,\dots, \mathcal B_4$ are the kinematical factors, \ref{B}, associated with each of the four scattering amplitudes $\mathcal{M}_1,\dots,\mathcal{M}_4$.\footnote{The terms $i=j$ do not contribute since $v_{ij}=0$.}

Next consider the case where the initial state contains $m$ electrons. The number of possible pairs of interacting particles is $N_{\rm pairs}=m(m-1)/2$. We get
\begin{equation} 
S_{ent,sing}=
-\frac{N_{\rm pairs}\,T\,\ln{e^6}}{32 \pi\, V}
\ln\left(\frac{E_d}{\lambda}\right)\,
\bigg\langle\,\frac{v_{ij}}{E_{ij}^2}\, \int_{0}^{\pi}d\theta \sin{\theta}\,
|i{\cal{M}}_{1}(E_{ij},\theta)|^2\,
\mathcal{B}_1({i'j';ij})\,\bigg\rangle
\end{equation} 
where
\begin{equation}
\bigg\langle\,\frac{v_{ij}}{E_{ij}^2}\, \int_{0}^{\pi}d\theta \sin{\theta}\,
\big|i{\cal{M}}_{1}(E_{ij},\theta)\big|^2\,
\mathcal{B}_1({i'j';ij})\,\bigg\rangle=
\frac{1}{N_{\rm pairs}}\,\sum_{i,j=1}^m\,\frac{v_{ij}}{2E_{ij}^2}\, \int_{0}^{\pi}d\theta \sin{\theta}\,
|i{\cal{M}}_{1}(E_{ij},\theta)|^2\,
\mathcal{B}_1({i'j';ij})
\end{equation}
is the two electron result of $\cite{TT}$, averaged over all possible interacting pair configurations.

The entanglement entropy per unit time per particle density is equal to
\begin{equation} 
s_{ent,sing}=\frac{VS_{ent,sing}}{Tm}
=
-\frac{(m-1)\,\ln{e^6}}{64 \pi}
\ln\left(\frac{E_d}{\lambda}\right)\,
\bigg\langle\,\frac{v_{ij}}{E_{ij}^2}\, \int_{0}^{\pi}d\theta \sin{\theta}\,
|i{\cal{M}}_{1}(E_{ij},\theta)|^2\,
\mathcal{B}_1({i'j';ij})\,\bigg\rangle
\end{equation} 
and diverges logarithmically in the continuum, $\lambda\to 0$ limit. Notice the scaling with the number of particles in the initial state, implying that the entanglement entropy is further enhanced as compared to the two-electron scattering case.

Of course the appearance of logarithmic divergences signals the breakdown of perturbation theory. Rather the perturbative analysis is valid when the size of the box is kept large, but fixed, taking the coupling $e$ to be sufficiently small. As shown in \cite{TT}, upon restricting the partial trace to be over the additional radiated soft photons, which energies in the range $E_d <\omega_\gamma <E$, and thus excluding the soft photons in the clouds, regulates the IR logarithmic divergences, and leads to well-behaved results in the continuum limit. This type of tracing alleviates the decoherence of the reduced density matrix \cite{Gomez1,Gomez2}.   

\subsection{Relation to the cusp anomalous dimension}
The integrand in \ref{entmaster} diverges for forward ($\theta=0$) and backward ($\theta=\pi$) scattering. Since forward or backward scattering cannot be distinguished from no scattering, we introduce an effective lower and upper cutoff on the scattering angle, $\theta_0\le \theta \le \pi -\theta_0$ where $\theta_0$ is a small angle, to regularize the integral. As a result, the dominant contributions to the leading entanglement entropy arise from the regions $\theta \simeq \theta_0$ and $\theta \simeq \pi- \theta_0$, close to the cutoffs. 

These dominant contributions exhibit certain universal behavior, irrespectively of the details of the initial state. To see this, recall that the squares of the amplitudes for the processes $e^-e^-\to e^-e^-$, $e^+e^+\to e^+e^+$ and $e^+e^-\to e^+e^-$ in the center of mass frame are given, respectively, by
$$
\frac{1}{4}\sum_{\rm spins}|i{\cal M}_1(E,\theta)|^2 ~ =~\frac{1}{4} \sum_{\rm spins}|i{\cal M}_2(E,\theta)|^2~ =~\frac{1}{4} \sum_{\rm spins}|i{\cal M}_3(E,\theta)|^2 
$$
\begin{equation}
=4 e^4 \left(\frac{E^2+p^2}{p^2}\right)^2 \left(\frac{4}{\sin^4\theta}-\frac{3}{\sin^2\theta} + \left(\frac{p^2}{E^2+p^2}\right)^2 \left(1+\frac{4}{\sin^2\theta}\right)\right)
\end{equation}
where $E$ and $p=|\vec{p}|$ are the energy and momentum of each one of the incoming particles, while $\theta$ is the scattering angle. As before, we have averaged over the polarizations of the initial particles for simplicity. The above three amplitudes diverge for forward and backward scattering. For the process $e^+e^- \to \gamma \gamma$, we obtain
\begin{equation}
\frac{1}{4}\sum_{\rm spins}|i{\cal M}_4(E,\theta)|^2 = 4 e^4 \left(\frac{E^2+p^2(1+\sin^2\theta)}{E^2-p^2\cos^2\theta} - \frac{2p^4\sin^4\theta}{(E^2-p^2\cos^2\theta)^2}\right)
\end{equation}
which remains finite and well-behaved at $\theta=0$ and $\theta=\pi$. As usual, it is useful to define the Mandelstam variables
\begin{equation}
s= 4E^2,\,\,\, t=-4p^2\sin^2{\frac{\theta}{2}}, \,\,\, u=-4p^2\cos^2{\frac{\theta}{2}}, \,\,\,\,\, s+t+u=4m^2
\end{equation}

We will examine the behavior of the dominant terms at $\theta_0$ and $\pi-\theta_0$ in a particular double scaling limit. We take the momentum to be very large and the cutoff angle very small, $p\to \infty$ and $\theta_0 \to 0$, keeping the quantity $\xi=4p^2\sin^2{(\theta_0/2)}$ fixed and large. When $\theta=\theta_0$, $u\simeq -s \to \infty$ in this limit, with $t=-\xi$ remaining fixed and large. Similarly, when $\theta=\pi-\theta_0$, $t\simeq -s \to \infty$, with $u=-\xi$ fixed. For the first three invariant amplitudes squared,
we obtain in this limit
\begin{equation}
\frac{1}{4}\sum_{\rm spins}|i{\cal M}_i(E,\theta_0)|^2=\frac{1}{4}\sum_{\rm spins}|i{\cal M}_i(E,\pi-\theta_0)|^2 \simeq 64 e^4 \frac{1}{\sin^4\theta_0} \simeq 64 e^4~ \frac{p^4}{\xi^2},\,\,\,\,\, i=1,2,3
\end{equation}
while
\begin{equation}
\frac{1}{4}\sum_{\rm spins}|i{\cal M}_4(E,\theta_0)|^2=\frac{1}{4}\sum_{\rm spins}|i{\cal M}_4(E,\pi-\theta_0)|^2 \simeq 8 e^4~ \frac{p^2}{m^2+\xi}
\end{equation}
Evidently, the amplitude squared for the process $e^+e^- \to \gamma \gamma$ is suppressed in this limit, as compared to the other three. So this contribution can be neglected.\footnote{The kinematical factor $\mathcal B_4$ grows logarithmically with the momentum $p$ in the limit, but not fast enough to alter this conclusion.}

The kinematical factors $\mathcal B_1$ and $\mathcal B_2$ appearing in \ref{entmaster} are given by
\begin{equation}
\mathcal B_1 = \mathcal B_2 ~=~\frac{e^2}{4\pi^2}~\bigg[~\frac{1-\frac{2m^2}{t}}{\sqrt{1-\frac{4m^2}{t}}}\ln\Bigg(\frac{1-\frac{2m^2}{t}+\sqrt{1-\frac{4m^2}{t}}}{1-\frac{2m^2}{t}-\sqrt{1-\frac{4m^2}{t}}}\Bigg)~+~\left(t \leftrightarrow u\right)~-~\left(t \leftrightarrow s\right)~-~2~\bigg]
\end{equation}
while $\mathcal B_3$ is given by
\begin{equation}
\mathcal B_3 = ~-\frac{e^2}{4\pi^2}~\bigg[~\frac{1-\frac{2m^2}{t}}{\sqrt{1-\frac{4m^2}{t}}}\ln\Bigg(\frac{1-\frac{2m^2}{t}+\sqrt{1-\frac{4m^2}{t}}}{1-\frac{2m^2}{t}-\sqrt{1-\frac{4m^2}{t}}}\Bigg)~+~\left(t \leftrightarrow u\right)~-~\left(t \leftrightarrow s\right)~+~2~\bigg]
\end{equation}
In the double scaling limit, we obtain at both $\theta_0$ and $\pi-\theta_0$ 
\begin{equation}
\mathcal B_1 = \mathcal B_2 =-\mathcal B_3=\frac{e^2}{4\pi^2}\ln\left(\frac{\xi}{4m^2}\right)
\end{equation}
This factor is precisely equal to the cusp anomalous dimension in QED, $e^2 \Gamma({\varphi})/4\pi^2$, via the relation $\xi=2m^2 (\cosh{\varphi} - 1)$, which controls the vacuum expectation value of a Wilson loop with a cusp of angle $\varphi$ \cite{Korchemsky,Sever2}.

Collecting all factors together, we arrive at a universal formula for the leading perturbative differential entanglement entropy, per unit time, per particle density, in the double scaling limit 
\begin{equation}
\frac{ds_{ent,sing}}{d\Omega}\large|_{\theta=\theta_0,\,\pi-\theta_0}=-\frac{e^6\ln{e^6}}{\pi^2\,N\,\sin^4{\theta_0}}\ln\left(\frac{E_d}{\lambda}\right)~\sum_{ij}~\frac{v_{ij}}{2E_{ij}^2}~ \hat{e}_i\, \hat{e}_j~ \Gamma({\varphi}_{ij})/4\pi^2 
\end{equation} 
where the sums are over all charged particles in the initial state; $\hat{e}_i$ is $+1$ for positrons and $-1$ for electrons, and the cusp angles $\varphi_{ij}$ are defined via the relation $\xi_{ij}=4p_i^2\sin^2{(\theta_0/2)}=2m^2 (\cosh{\varphi_{ij}} - 1)$. Thus, the coefficient of the logarithmic singularity is given in terms of the cusp anomalous dimension of QED, averaged over all pairs of interacting particles via the above equation.


\section{Next to leading order analysis for electron/electron scattering} \label{s3}
\setcounter{equation}{0}
In this section we extend the perturbative analysis to the next to leading order and obtain ${\cal{O}}(e^8)$ corrections to the Renyi and entanglement entropies. For simplicity, we focus on a Fock basis computation concerning the case of two-electron scattering. Thus we set the dressing function $f^{\mu}$ equal to zero. As we have seen from the leading order analysis, the singular, logarithmically divergent part of the Renyi and entanglement entropies is independent of the dressing. 

\subsection{The large eigenvalue of $\rho_H$ and the Renyi entropies to order $e^8$}

Two-photon outgoing states, $\ket{\beta\gamma_1\gamma_2}$, contribute to next to leading order and must be properly taken into account. For such a state to be in the soft part of the Hilbert space ${\cal{H}}_S$, the sum of the energies $E_{\gamma_1}+E_{\gamma_2}$ must be below the reference energy scale $E$. Therefore, even if both $E_{\gamma_1}<E$ and $E_{\gamma_2}<E$, if $E_{\gamma_1}+E_{\gamma_2}>E$, the two-photon state is taken to be in ${\cal{H}}_H$. If $E_{\gamma_1}>E$ and  $E_{\gamma_2}<E$, then $\ket{\beta\gamma_1\gamma_2}\to \ket{\beta\gamma_1}_H\times \ket{\gamma_2}_S$. 
 
The outgoing state is given by
\begin{equation}
\ket{\Psi}_{out}=S\ket{\alpha}=\ket{\alpha}+\sum_{\beta}T_{\beta\alpha}\ket{\beta}+\sum_{\beta,\gamma}T_{\beta\gamma,\alpha}\ket{\beta\gamma}+\sum_{\substack{\beta,\gamma_1,\gamma_2\\\omega_{\gamma_1}\leq\omega_{\gamma_2}}}T_{\beta\gamma_1\gamma_2,\alpha}\ket{\beta\gamma_1\gamma_2}+\dots
\end{equation}
where $T_{\beta\gamma_1\gamma_2,\alpha}=\bra{\beta\gamma_1\gamma_2}iT\ket{\alpha}$ is the amplitude for the two incoming electrons (described by the state $\ket{\alpha}$) to scatter and produce the electron/positron state $\ket{\beta}$, emitting at the same time two photons ($\ket{\gamma_1\gamma_2}$). The set $\{\ket{\beta}\}$ comprises of electron/positron states with zero number of photons. To order $e^8$, we must consider states with two electrons and states with two electrons and an electron/positron pair.\footnote{As it turns out, the latter do not contribute to the Renyi and entanglement entropies to this order.} The amplitudes $T_{\beta\alpha}$ are of order $e^2$ for the two-electron case and $e^4$ for the case of three electrons and a positron. Likewise, the amplitudes $T_{\beta\gamma,\alpha}$, describing the emission of an additional photon, are of order $e^3$ and $e^5$, respectively. Finally, $T_{\beta\gamma_1\gamma_2,\alpha}$ is of order $e^4$ if $\ket{\beta}$ is a two-electron state and of order $e^6$ if $\ket{\beta}$ is a state of three electrons and a positron. States with more than two photons will not affect the entanglement and Renyi entropies to order $e^8$. So we do not write them explicitly in the equation above.

The reduced density matrix is obtained upon taking a partial trace over ${\cal{H}}_S$: $\rho_H=\Tr_{H_S}\ket{\Psi}_{out}\bra{\Psi}_{out}$. We define the state 
\begin{equation}
\ket{\Phi}=\ket{\alpha}_H+ \sum_{\beta}T_{\beta\alpha}\ket{\beta}_H+\sum_{\beta}\sum_{\omega_\gamma>E}T_{\beta\gamma,\alpha}\ket{\beta\gamma}_H+\sum_\beta\sum_{\ket{\gamma_1\gamma_2}\in {\cal{H}}_H}T_{\beta\gamma_1\gamma_2,\alpha}\ket{\beta\gamma_1\gamma_2}_H+\dots
\end{equation}
and write the reduced density matrix as
\begin{equation}
\rho_H=\ket{\Phi}\bra{\Phi} +G
\end{equation}
where to order $e^8$ $G$ is given by
$$
G=\bigg(\sum_{\omega_\gamma<E}T_{\beta\gamma,\alpha}T^*_{\beta'\gamma,\alpha}+\sum_{\ket{\gamma_1\gamma_2}\in {\cal{H}}_S}T_{\beta\gamma_1\gamma_2,\alpha}T^*_{\beta'\gamma_1\gamma_2,\alpha}\bigg)\ket{\beta}_H\bra{\beta'}_H 
$$
\begin{equation}
+ \bigg(\sum_{\substack{\omega_{\gamma}<E}}T_{\beta\gamma,\alpha}T^*_{\beta'\gamma\gamma',\alpha}\ket{\beta}_H\bra{\beta'\gamma'}_H + h.c.\bigg) + \sum_{\substack{\omega_{\gamma_1}<E}}T_{\beta\gamma\gamma_1,\alpha}T^*_{\beta'\gamma'\gamma_1,\alpha}\ket{\beta\gamma}_H\bra{\beta'\gamma'}_H 
\end{equation}
The first sum in the first line includes cross-terms between two-electron states and states with three electrons and a positron.

Using the fact that at finite $\lambda$, $T^*_{\alpha\gamma,\alpha}=T^*_{\alpha\gamma_1\gamma_2,\alpha}=0$ by energy conservation and $\sum_{\beta}T_{\beta\alpha}T^*_{\beta\gamma,\alpha}=0$ by incompatibility of the corresponding energy-momentum $\delta$-functions, we see that $G$ annihilates $\ket{\Phi}$ to order $e^8$. So $\ket{\Phi}$ is an eigenstate of $\rho_H$ with eigenvalue $\bra{\Phi}\ket{\Phi}$. Thus to this order there is a ``large'' eigenvalue given by 
\begin{equation}
\lambda_\Phi=\bra{\Phi}\ket{\Phi}=1+ T_{\alpha\alpha}+T^*_{\alpha\alpha}+\sum_{\beta}T_{\beta\alpha}T^*_{\beta\alpha}+\sum_{\beta}\sum_{\omega_\gamma>E}T_{\beta\gamma,\alpha}T^*_{\beta\gamma,\alpha}+\sum_\beta\sum_{\ket{\gamma_1\gamma_2}\in {\cal{H}}_H}T_{\beta\gamma_1\gamma_2,\alpha}T^*_{\beta\gamma_1\gamma_2,\alpha}
\end{equation}
Using unitarity \ref{unitarity}, the eigenvalue can be written as follows
\begin{equation}
\lambda_\Phi=1-\Delta
\end{equation}
where to order $e^8$ 
\begin{equation}
\Delta=\sum_{\beta}\sum_{\omega_\gamma<E}T_{\beta\gamma,\alpha}T^*_{\beta\gamma,\alpha}+\sum_\beta\sum_{\ket{\gamma_1\gamma_2}\in {\cal{H}}_S}T_{\beta\gamma_1\gamma_2,\alpha}T^*_{\beta\gamma_1\gamma_2,\alpha} + \sum_{\beta}\sum_{\omega_{\gamma_1}<E,\omega_{\gamma_2}>E}T_{\beta\gamma_1\gamma_2,\alpha}T^*_{\beta\gamma_1\gamma_2,\alpha} \label{Delta8}
\end{equation}
States with three electrons and a positron do not contribute to $\Delta$ to order $e^8$, and so the sums over $\beta$ in the expression above can be restricted to two-electron states. 

The rest of the non-zero eigenvalues of $\rho_H$ coincide with the non-zero eigenvalues of $G$. These split into two groups, the eigenvalues of order $e^6$ (which may receive corrections of order $e^8$) and the eigenvalues of order $e^8$. The sum of all of these must be equal to $\Delta$, as can be verified by computing $\Tr G$.

The traces $\Tr (\rho_H)^m$ for integer $m\ge 2$ are given by
\begin{equation}\label{tracerho}
\Tr (\rho_H)^m=1-m\Delta
\end{equation}
This equation can be verified by direct computation of the traces to order $e^8$ -- see \ref{a4}.
The corresponding Renyi entropies take the form
\begin{equation}   
S_{m}=-\frac{1}{m-1}\ln(1-m\Delta)=\frac{m}{m-1}\Delta   
\end{equation}

Notice that to all orders, the traces satisfy the inequalities $0<\Tr (\rho_H)^m\le 1$. So the perturbative result \ref{tracerho} breaks down at large $m \sim 1/\Delta$. 
To regulate the perturbative expansion of the entanglement entropy, we consider instead the generating functional \cite{DHoker}
\begin{equation}\label{functional}
G(w, \rho_H)=-\Tr \left(\rho_H \ln[1-w(1-\rho_H)]\right) 
\end{equation}
The entanglement entropy is recovered in the limit $w\to 1$
\begin{equation}
S_{ent}=\lim_{w\to 1}G(w, \rho_H)
\end{equation}
 The Taylor expansion of $G(w, \rho_H)$ is convergent for $|w|<1$ and gives
 \begin{equation}\label{functionaltaylor}
 G(w, \rho_H)=\sum_{n=1}^{\infty}\,\frac{w^n}{n}\,\left[\sum_{m=0}^n \frac{n!}{(n-m)!m!}\,(-1)^m\,\Tr(\rho_H)^{m+1}\right]
 \end{equation}
In fact, setting $w=e^{-\kappa}$, for $\kappa \sim \Delta$, suppresses the contributions of the traces for large integer $m\sim 1/\Delta$ and regulates the perturbative expansion.
Using \ref{tracerho}, we obtain to order $e^8$ at fixed $w$
\begin{equation}
G(w, \rho_H)=\Delta(-\ln(1-w)+1)=-\Delta \ln{\kappa} + \Delta
\end{equation}
The leading perturbative entanglement entropy computed in the previous section is recovered to be $-\Delta \ln{e^6}$. Because of the logarithmic cutoff term, there are  ambiguities in the analytic, cutoff independent part. The perturbative expansion does not commute with the  $w\to 1$ limit.

\subsection{Entanglement entropy to order $e^8$}
We proceed now to discuss next to leading order corrections to the entanglement entropy. In particular, we will study the structure of the non-analytic part (in the electron charge $e$) of the entanglement entropy, which depends on the non-vanishing, ``small" eigenvalues of $\rho_H$. As we remarked in the previous section, these eigenvalues coincide with the (non-zero) eigenvalues of $G$. They split into two groups, of order $e^6$ and $e^8$, respectively, and their sum is given by $\sum_{i\ne \Phi}\lambda_i =1-\lambda_\Phi=\Delta$. Let $\Delta_6$ and $\Delta_8$ be the order $e^6$ and the order $e^8$ parts of $\Delta$, respectively.

In \ref{a5}, we describe how to obtain the shifts of the order $e^6$ eigenvalues of $G$ using second order perturbation theory.
Let us denote the sum of these shifts with $\tilde \Delta_8$. 
It is then clear that the sum of the order $e^6$ eigenvalues of $G$ is given by $\Delta_6+\tilde \Delta_8$, while the sum of all the order $e^8$ eigenvalues by $\Delta - (\Delta_6+\tilde \Delta_8)=\Delta_8-\tilde \Delta_8$. These sums determine the non-analytic part of the entanglement entropy
\begin{equation}
S_{ent}=-\sum_i \lambda_i \ln{\lambda_i}=-(\Delta_6+\tilde \Delta_8)\ln e^6 -(\Delta_8-\tilde \Delta_8)\ln e^8 ~+~\rm{analytic~ in~ e}
\end{equation}
Thus the non-analytic part is given by
\begin{equation}
(S_{ent})_{N.A.}=-\bigg(\Delta_6+\frac{4\Delta_8}{3}-\frac{\tilde \Delta_8}{3}\bigg)\ln e^6
\end{equation}

\subsection{IR divergences to order $e^8$}
The IR divergent terms in the Renyi and entanglement entropies can be obtained by examining the singular part of $\Delta$ (which gives the shift of the ``large'' eigenvalue from unity). When the energy scale $E$ is sufficiently small, we can apply soft theorems for real photon emission to obtain (in the continuum limit) \cite{Weinberg}
\begin{equation}
\Delta = \sum_{\beta} T_{\beta\alpha}T^*_{\beta\alpha} \left[2\mathcal{B_{\beta\alpha}}\ln(\frac{E}{\lambda}) + 2 \left(\mathcal{B_{\beta\alpha}}\ln(\frac{E}{\lambda})\right)^2\right]+ \sum_{\beta}\sum_{\omega_\gamma>E}T_{\beta\gamma,\alpha}T^*_{\beta\gamma,\alpha}\, \left(2\mathcal{B_{\beta\gamma,\alpha}}\ln(\frac{E}{\lambda})\right)
\end{equation}

Now this expression must be computed to order $e^8$. So the amplitude squared in the first term should be calculated to order $e^6$, and, therefore one-loop diagrams contribute. As a result, logarithmic IR divergences due to virtual photons appear. Soft theorems for virtual photons give \cite{Weinberg} 
\begin{equation}
T_{\beta\alpha}T^*_{\beta\alpha}=(T_{\beta\alpha}T^*_{\beta\alpha})^\Lambda\left(\frac{\lambda}{\Lambda}\right)^{2\mathcal{B_{\beta\alpha}}}=(T_{\beta\alpha}T^*_{\beta\alpha})^\Lambda \left(1-2\mathcal{B_{\beta\alpha}}\ln(\frac{E}{\lambda}) - 2\mathcal{B_{\beta\alpha}}\ln(\frac{\Lambda}{E})\right)+\dots
\end{equation}  
where scattering amplitude squared  $(T_{\beta\alpha}T^*_{\beta\alpha})^\Lambda$ does not include contributions from virtual photons with energy below the cutoff scale $\Lambda \sim E$. The factor $\ln(\Lambda/E)$ is negligible when the reference scales $E$ and $\Lambda$ are of the same order. Substituting into the expression for $\Delta$, yields   
\begin{equation}
\Delta = \sum_{\beta} (T_{\beta\alpha}T^*_{\beta\alpha})^\Lambda \left[2\mathcal{B_{\beta\alpha}}\ln(\frac{E}{\lambda}) - 2 \left(\mathcal{B_{\beta\alpha}}\ln(\frac{E}{\lambda})\right)^2\right]+ \sum_{\beta}\sum_{\omega_\gamma>E}T_{\beta\gamma,\alpha}T^*_{\beta\gamma,\alpha}\, \left(2\mathcal{B_{\beta\gamma,\alpha}}\ln(\frac{E}{\lambda})\right)
\end{equation}

The first term is in accordance with an exponentiation pattern. Notice however the appearance of the last term which originates from two-photon final states in which one photon is soft and the other is hard. The logarithmic divergence is now proportional to the amplitude squared for real (hard) photon emission, and a different kinematical factor. So the exponentiation pattern of the IR logarithmic divergences is an intricate one.


\section{To all orders entanglement in electron/electron scattering}\label{s4}
\setcounter{equation}{0}
In this section we attempt to generalize some of the results of the previous sections to all orders in perturbation theory so as to gain insight into the exponentiation pattern of the IR logarithmic divergences, and to link our perturbative results with the results of \cite{Carney1,Carney2,Carney3,Carney4} (see also \cite{Gomez1,Gomez2}) on the decoherence of the density matrix $\rho_H$. As in \ref{s3}, we focus on a Fock basis computation associated with two-electron scattering for simplicity.

We denote by $\ket{{\bm \gamma}}=\ket{\gamma_1\gamma_2,\dots,\gamma_n}$ a generic multiphoton box state. The number of photons $n$ is nonzero and finite, but arbitrary. It will be convenient to order the photons in ascending order, according to their energy: $E_1\le E_2\le\dots\le E_n$. If the total energy of the photons is less that the reference energy scale $E$, $\sum_{i=1}^n E_i < E$, then $\ket{{\bm \gamma}}\in {\cal{H}}_S$. Suppose $\sum_{i=1}^n E_i > E$. In order to decompose $\ket{{\bm \gamma}}$ in ${\cal{H}}_H\times {\cal{H}}_S$, we first determine the maximum subset of photons $\{\gamma_1,\dots\gamma_m\}$, for some positive integer $m$, each having energy below the reference scale $E$, so that $\sum_{i=1}^m E_{\gamma_i}<E$ {\it and} $\sum_{i=m+1}^n E_{\gamma_i}\ge E$. Then we write the multiphoton state as a product state as follows 
\begin{equation}
\ket{{\bm \gamma}}=\ket{\gamma_1\gamma_2,\dots,\gamma_n}=\ket{\gamma_{m+1},\gamma_{m+2},\dots,\gamma_n}_H\times \ket{\gamma_1\gamma_2,\dots,\gamma_m}_S
\end{equation} 
If no such $m$ exists, then the state $\ket{{\bm \gamma}}$ is in ${\cal{H}}_H$. Photons with energy scaling with the IR cutoff $\lambda$ will appear in the soft part of the Hilbert space in the continuum, $\lambda \to 0$ limit.\footnote{Alternatively, we could take any photon with energy below $E$ to lie in ${\cal{H}}_S$, without imposing any restriction on the total energy of the soft and hard factors of the state.}       

Adopting these notations, the outgoing state can be written as
\begin{equation}
\ket{\Psi}_{out}=S\ket{\alpha}=\ket{\alpha}+\sum_{\beta}T_{\beta\alpha}\ket{\beta}+\sum_{\ket{\beta{\bm\gamma}}}T_{\beta{\bm\gamma},\alpha}{\ket{\beta{\bm\gamma}}}
\end{equation}
at any order in perturbation theory. The set $\{\ket{\beta}\}$ comprises of states with $2+l$ electrons, $l$ positrons states and zero number of photons, with $l=0,1,\dots$. The reduced density matrix is given by $\rho_H=\Tr_{H_S}\ket{\Psi}_{out}\bra{\Psi}_{out}$. As before, we define 
\begin{equation}
\ket{\Phi}=\ket{\alpha}_H+ \sum_{\beta}T_{\beta\alpha}\ket{\beta}_H+\sum_{\beta}\sum_{\ket{\bm \gamma}\in {\cal{H}}_H}T_{\beta{\bm\gamma},\alpha}\ket{\beta{\bm\gamma}}_H
\end{equation}
in terms of which the reduced density matrix is given by
\begin{equation}
\rho_H=\ket{\Phi}\bra{\Phi} +G
\end{equation}
where
$$
G=\sum_{\ket{\bm \gamma'}\in {\cal{H}}_S}T_{\beta{\bm \gamma'},\alpha}T^*_{\beta'{\bm \gamma'},\alpha}~\ket{\beta}_H\bra{\beta'}_H + \bigg(\sum_{\ket{\bm \gamma'}\in {\cal{H}}_S}T_{\beta{\bm \gamma'},\alpha}T^*_{\beta'{\bm \gamma}{\bm \gamma'},\alpha}\ket{\beta}_H\bra{\beta'{\bm \gamma}}_H + h.c.\bigg)$$
\begin{equation}
 + \sum_{\ket{\bm \gamma''}\in {\cal{H}}_S}T_{\beta{\bm \gamma}{\bm \gamma''},\alpha}T^*_{\beta'{\bm \gamma'}{\bm \gamma''},\alpha}\ket{\beta{\bm \gamma}}_H\bra{\beta'{\bm \gamma'}}_H 
\end{equation}
The expression for $G$ is valid to all orders in perturbation theory.

Next we impose energy-momentum conservation at finite $\lambda$: for any $\ket{\bm \gamma'},\, \ket{\bm \gamma''} \in {\cal{H}}_S$, $T^*_{\alpha{\bm\gamma'},\alpha}=0$ and $\sum_\beta T_{\beta\alpha}T^*_{\beta{\bm \gamma'},\alpha}=0$, $\sum_{\beta}\sum_{\ket{\bm \gamma}\in {\cal{H}}_H}T_{\beta{\bm \gamma},\alpha}T^*_{\beta{\bm\gamma}{\bm \gamma''},\alpha}=0$ due to the incompatibility of the corresponding energy-momentum $\delta$-functions. As a result $G$ annihilates $\ket{\Phi}$. We conclude that $\ket{\Phi}$ is an exact eigenstate of the reduced density matrix with eigenvalue $\lambda_{\Phi}=\bra{\Phi}\ket{\Phi}$.

Explicitly $\lambda_\Phi$ is given by
\begin{equation}
\lambda_{\Phi}=1+ T_{\alpha\alpha}+T^*_{\alpha\alpha}+\sum_{\beta}T_{\beta\alpha}T^*_{\beta\alpha}+\sum_{\beta}\sum_{\ket{\bm \gamma}\in {\cal{H}}_H}T_{\beta{\bm\gamma},\alpha}T^*_{\beta{\bm \gamma},\alpha}
\end{equation}
The sum of the rest of the eigenvalues of $\rho_H$ is given by
\begin{equation}
\Delta=\sum_{i\neq \Phi}\lambda_i =1-\lambda_{\Phi}=-\bigg(T_{\alpha\alpha}+T^*_{\alpha\alpha}+\sum_{\beta}T_{\beta\alpha}T^*_{\beta\alpha}+\sum_{\beta}\sum_{\ket{\bm \gamma}\in {\cal{H}}_H}T_{\beta{\bm\gamma},\alpha}T^*_{\beta{\bm \gamma},\alpha}\bigg)
\end{equation}

We remark that the amplitudes $T_{\beta\alpha}, T_{\beta{\bm\gamma},\alpha}$ appearing in the above expressions for $\Delta$ and $\lambda_{\Phi}$ are hard, in that the final state does not include soft photons with energies scaling with $\lambda$ in the continuum limit. At any finite order in perturbation theory, these amplitudes are plagued by IR logarithmic divergences, due to virtual soft photons running in the loops \cite{Weinberg}, when the momenta of the two-electron states $\alpha$ and $\beta$ differ. Since the corresponding transition processes do not involve the emission of soft radiation, there are no similar divergent contributions from real soft photons to $\lambda_\Phi$ (and $\Delta$) to cancel the virtual IR divergences. 

However, the term $T_{\alpha\alpha}+T^*_{\alpha\alpha}$ is free of any IR divergences, order by order in perturbation theory, since this is related to the {\it total inclusive cross-section} in the state $\alpha$ by unitarity, \ref{unitarity}:
\begin{equation}
T_{\alpha\alpha}+T^*_{\alpha\alpha}=-\sum_{\beta}T_{\beta\alpha}T^*_{\beta\alpha}-\sum_{\beta}\sum_{\ket{\bm \gamma}}T_{\beta{\bm\gamma},\alpha}T^*_{\beta{\bm \gamma},\alpha}\label{imaginary}
\end{equation}
The corresponding transition processes include the emission of real soft photons. As is well known, such inclusive transition rates are IR finite, with the IR divergences due to virtual soft photons cancelling against the ones due to real photon emission \cite{Weinberg}. We emphasise that $T_{\alpha\alpha}$ is a {\it box transition amplitude}, scaling inversely proportional with the volume of the box in the $\lambda \to 0$, continuum limit:
\begin{equation}
T_{\alpha\alpha}\to \frac{T}{VE_{12}^2} \, i{\cal{M}}_{\alpha\alpha}
\end{equation} 
where $T$ is the time-scale of the experiment, $i{\cal{M}}_{\alpha\alpha}$ is the invariant amplitude at infinite volume \footnote{We follow the normalizations of \cite{Peskin} so that the invariant amplitude $i{\cal{M}}_{\beta\alpha}$ has units of length to the power $N_f-2$, where $N_f$ is the number of particles in the final state $\beta$.} and $E_{12}$ is the total energy in the center of mass frame of the two electrons of the state $\ket{\alpha}$. In particular, 
\begin{equation}
T_{\alpha\alpha}+T^*_{\alpha\alpha}=-\frac{T}{VE_{12}^2} \, 2{\rm Im}{\cal{M}}_{\alpha\alpha}=-\frac{Tv_{ij}}{V}\,\Sigma_\alpha
\end{equation}  
where $v_{ij}$ is the relative velocity of the two electrons in the state $\alpha$ and $\Sigma_\alpha$ is the total, inclusive cross-section in this state.

As a result, at any finite order in perturbation theory IR logarithmic divergences appear in the expressions for the Renyi and the entanglement entropies, as we have seen explicitly at leading and next-to-leading order in \ref{s2} and \ref{s3}.  

To all orders, however, the virtual IR logarithmic divergences exponentiate \cite{Weinberg}
\begin{equation}
T_{\beta\alpha}T^*_{\beta\alpha}=(T_{\beta\alpha}T^*_{\beta\alpha})^\Lambda\left(\frac{\lambda}{\Lambda}\right)^{2\mathcal{B_{\beta\alpha}}}, \,\,\, T_{\beta{\bm\gamma},\alpha}T^*_{\beta{\bm \gamma},\alpha}=(T_{\beta{\bm \gamma},\alpha}T^*_{\beta{\bm \gamma},\alpha})^\Lambda\left(\frac{\lambda}{\Lambda}\right)^{2\mathcal{B_{\beta\alpha}}}
\end{equation}
leading to the vanishing of the hard amplitudes $T_{\beta\alpha}, T_{\beta{\bm \gamma},\alpha}$ when the momenta of $\beta$ and $\alpha$ differ. The vanishing of these hard amplitudes can be also understood as a consequence of symmetry \cite{StromingerLectures,StromingerIRrevisited}. Taking into account the scaling of the box amplitude $T_{\alpha\alpha}$ in the large volume limit, we conclude that {\it to all orders}, $\lambda_{\Phi}$ becomes equal to a diagonal element of $\rho_H$ 
\begin{equation}
\lambda_{\Phi} = 1 + T_{\alpha\alpha}+T^*_{\alpha\alpha} = 1 - \frac{Tv_{ij}}{V}\,\Sigma_\alpha
\end{equation}
and 
\begin{equation}
\Delta= \frac{Tv_{ij}}{V}\,\Sigma_\alpha
\end{equation}
{\it free of any IR divergences in $\lambda$.} Indeed, to all orders, the reduced density matrix $\rho_H$ assumes a diagonal form in terms of the momentum indices exhibiting decoherence \cite{Carney1,Carney2,Carney3,Carney4}. The diagonal elements (in momentum) are given in terms of inclusive Bloch-Nordsieck rates associated with box states, and are free of IR divergences. 

Assuming that $\Sigma_{\alpha}$ is finite, we see that to all orders, the reduced density matrix is dominated by a large eigenvalue, $\lambda_{\Phi}$, in the continuum, $\lambda\to 0$ limit. The rest of the eigenvalues must be small, scaling inversely proportional with powers of the volume of the box. Despite the fact that the number of the small eigenvalues also grows with the volume,
the traces $\Tr (\rho_H)^m$ for integer $m\ge 2$ are given by
\begin{equation}\label{Tracesall}
\Tr (\rho_H)^m=1-m\Delta=1-m\,\frac{Tv_{ij}}{V}\,\Sigma_\alpha
\end{equation}
and the corresponding Renyi entropies take the form
\begin{equation}\label{Renyiall}   
S_{m}=-\frac{1}{m-1}\ln(1-m\Delta)=\frac{m}{m-1}\,\frac{Tv_{ij}}{V}\,\Sigma_\alpha  
\end{equation}
The Renyi entropies per particle flux, per unit time
\begin{equation}   
s_{m}=\frac{m}{(m-1)}\,\Sigma_\alpha  
\end{equation}
remain finite in the limit. They are proportional to the total cross-section $\Sigma_{\alpha}$, free of IR divergences.

Let us discuss the dependence of $\Sigma_\alpha$ on the reference scale $E$, taking into account the scaling of the box amplitudes with the volume and the vanishing of the purely hard amplitudes to all orders in the continuum limit. \ref{imaginary} gives
\begin{equation}
\frac{Tv_{ij}}{V}\,\Sigma_\alpha  =-T_{\alpha\alpha}-T^*_{\alpha\alpha}=\sum_{\beta}~\sum_{\ket{\bm \gamma'}\in {\cal{H}}_S}~T_{\beta{\bm \gamma'},\alpha}T^*_{\beta{\bm \gamma'},\alpha}+\sum_{\beta}\sum_{\ket{\bm \gamma}\in {\cal{H}}_H}~\sum_{\ket{\bm \gamma'}\in {\cal{H}}_S}~T_{\beta{\bm\gamma}{\bm \gamma'},\alpha}T^*_{\beta{\bm \gamma}{\bm \gamma'},\alpha}
\end{equation}  
The total cross-section $\Sigma_\alpha$ is proportional to a sum of inclusive Bloch-Nordsieck rates, of the form,
\begin{equation}
\sum_{\ket{\bm \gamma'}\in {\cal{H}}_S}~T_{\beta{\bm\gamma}{\bm \gamma'},\alpha}T^*_{\beta{\bm \gamma}{\bm \gamma'},\alpha}\label{inclusive}
\end{equation} 
for the initial two electrons $\alpha$ to scatter and produce the hard state $\beta{\bm \gamma}$, emitting at the same time {\it any} number of soft photons with total energy less than $E$. Notice that in the continuum limit, the sums over $\beta$ and ${\bm \gamma}$ give rise to multiple integrals over the momenta of the hard final particles with appropriate volume factors (as well as discrete sums over the polarization indices) -- see \ref{cl}. Each such contribution scales proportionally with $T/V$ in the continuum limit, irrespectively of the number of particles in the final state.  

The inclusive rates \ref{inclusive} have been computed to all orders in $\cite{Weinberg}$, taking into account the cancellation of the IR divergences due to virtual soft photons and real soft photon emission
\begin{equation}
\sum_{\ket{\bm \gamma'}\in {\cal{H}}_S}~T_{\beta{\bm\gamma}{\bm \gamma'},\alpha}T^*_{\beta{\bm \gamma}{\bm \gamma'},\alpha}~=~(T_{\beta{\bm \gamma},\alpha}T^*_{\beta{\bm \gamma},\alpha})^\Lambda~b_{\beta\alpha}~\left(\frac{E}{\Lambda}\right)^{2\mathcal{B_{\beta\alpha}}} \label{inclusive1}
\end{equation}
where $b_{\beta\alpha}\simeq1-\pi^2\mathcal{B_{\beta\alpha}}^2/3$ \cite{Weinberg}. Therefore, we obtain
\begin{equation}
S_m\sim \frac{Tv_{ij}}{V}\,\Sigma_\alpha = \sum_{\beta}~(T_{\beta\alpha}T^*_{\beta\alpha})^\Lambda~b_{\beta\alpha}~\left(\frac{E}{\Lambda}\right)^{2\mathcal{B_{\beta\alpha}}} +\sum_{\beta}\sum_{\ket{\bm \gamma}\in {\cal{H}}_H}~(T_{\beta{\bm \gamma},\alpha}T^*_{\beta{\bm \gamma},\alpha})^\Lambda~b_{\beta\alpha}~\left(\frac{E}{\Lambda}\right)^{2\mathcal{B_{\beta\alpha}}} 
\end{equation}  
The ratio $E/\Lambda$ remains fixed in the $\lambda \to 0$ limit. To all orders, the small parameter controlling the expansion of the Renyi entropies in the continuum limit is set by the inverse of the product of the size of the box $L$ and the center of mass energy of the initial state.  
 
The entanglement entropy is not analytic in the volume of the box. (More precisely it is not analytic in the small dimensionless parameter discussed above). The non-analytic part is governed by the small eigenvalues, via the expressions $-\lambda_i\ln\lambda_i$, for $i \ne \Phi$. Ignoring the spin polarization structure, we may approximate the small eigenvalues with the diagonal elements of $\rho_H$, which are given in terms of inclusive rates   
\begin{equation}
D_{\beta{\bm \gamma},\beta{\bm\gamma}}=\sum_{\ket{\bm \gamma'}\in {\cal{H}}_S}T_{\beta{\bm \gamma}{\bm\gamma'},\alpha}T^*_{\beta{\bm \gamma}{\bm\gamma'},\alpha}
\end{equation}
where $\ket{\bm \gamma}$ is a hard photon state. This diagonal element scales with $T^2/V^{N_f}$ in the large volume limit, where $N_f$ is the number of particles in the final state, taking into account the energy-momentum conserving $\delta$-functions. It gives a contribution $-D_{\beta{\bm \gamma},\beta{\bm \gamma}}\ln (D_{\beta{\bm \gamma},\beta{\bm \gamma}})$ to the entanglement entropy. So terms logarithmic divergences in the size of the box (or the IR cutoff $\lambda$), of the form $\ln (V^{N_f}/T^2)$, in the expression for the entanglement entropy per particle flux per unit time persist to all orders in the continuum, large volume limit.

One way to recover the entanglement entropy is to take the $w\to 1$ limit of the generating functional $G(w, \rho_H)$, \ref{functional}, whose Taylor expansion is given in terms of the traces $\Tr(\rho_H)^m$ (and hence the exponentials of the Renyi entropies) -- see \ref{functionaltaylor}. How is it then that the all orders entanglement entropy (per particle flux, per unit time) retains non-analytic, logarithmic behavior with respect to the size of the box or the IR cutoff? Notice that for any given large $m$, the scaling of $\Tr(\rho_H)^m$ with the volume, \ref{Tracesall} (consequently the scaling of the corresponding Renyi entropy, \ref{Renyiall}), is valid for sufficiently large volume. At large but fixed volume, however, this scaling breaks down for large enough integers $m > 1/\Delta$, since the traces satisfy the strict inequalities $0<\Tr (\rho_H)^m\le 1$. A similar behavior was observed in the case of the perturbative results of the previous sections (see the discussion around \ref{functional}), leading to the non-analyticity of the entanglement entropy in the electron charge $e$. In the case at hand, the expansion parameter scales inversely proportional with the size of the box $L$. As a result, the $w\to 1$ limit {\it does not commute} with the large $L$ expansion of the traces. Indeed when $|w| < 1$, the contribution of the large $m$ traces in \ref{functionaltaylor} is suppressed. Using \ref{Tracesall} and the Taylor expansion of the generating functional, we obtain at large but fixed volume  
\begin{equation}
G(w,\rho_H)=(-\ln(1-w)+1)\Delta=(-\ln(1-w)+1)\, \frac{Tv_{ij}}{V}\,\Sigma_\alpha 
\end{equation}
However, this yields a diverging result in the limit $w\to 1$, indicating that we cannot reverse the order of limits. Similarly, another way to recover the entanglement entropy is to analytically continue the expression for the Renyi entropies for integer $m\geq 2$ to general real values, and take the limit $S_{ent}=\lim_{m\to 1}S_m$. However, this limit does not commute with the large volume expansion of the Renyi entropies, as can be seen from \ref{Renyiall}. Essentially, the entanglement entropy is non-analytic in the eigenvalues of the reduced density matrix. Since to all orders the small eigenvalues scale inversely proportional with the volume of the box, the entanglement entropy retains non-analytic behavior in the inverse of the volume or the IR cutoff.


\section{Conclusions}\label{s5}
\setcounter{equation}{0}
In this work we study generic scattering processes in QED in order to establish measures of the entanglement between the soft and hard particles in the final state. The initial state consists of an arbitrary number of Faddeev-Kulish electrons and positrons. To regulate the computation of the entanglement and the Renyi entropies, we place the system in a large box of size $L$, which provides an infrared momentum cutoff $\lambda \sim 1/L$. The perturbative computations are carried out at fixed $\lambda$, for sufficiently small coupling $e$. The density matrix for the hard particles is constructed via tracing over the entire spectrum of soft photons, including those in the clouds dressing the asymptotic charged particles. The leading perturbative behavior of the Renyi and entanglement entropies is governed by a large eigenvalue, which we compute at leading order, and next to leading order for two-electron scattering, in perturbation theory. The leading entanglement entropy and Renyi entropies for integer $m\ge 2$ are logarithmically divergent with respect to the IR cutoff $\lambda$ for all scattering cases we consider. We find that the coefficient of the logarithmic divergence is independent of the particle dressing and exhibits certain universality properties, irrespectively of the detailed properties of the initial state. The dominant contributions arise from two-particle interactions at forward and backward scattering. In the relativistic limit, these dominant contributions are proportional to the cusp anomalous dimension in QED. We see that the entanglement entropy in momentum space can be IR divergent, with the coefficient of the divergence containing physical information. We also study next to leading order corrections to the non-analytic part (in the electron charge $e$) of the entanglement entropy, and comment on the pattern of the exponentiation of the IR logarithmic divergences.

Since the perturbative expansion does not commute with the continuum, $\lambda \to 0$ limit due to the appearance of IR divergences, we proceed to study the behavior of the Renyi and entanglement entropies to all orders in perturbation theory. We focus on Fock basis computation associated with two-electron scattering for simplicity. We derive an exact expression for the large eigenvalue of the density matrix in terms of hard scattering amplitudes, which is valid at any finite order in perturbation theory. At any finite order in the electron charge $e$, this eigenvalue is plagued with IR logarithmic divergences due to soft virtual photons. To all orders in perturbation theory, the IR divergences exponentiate, leading to a finite result. In the large volume limit, the shift of this eigenvalue from unity governs the behavior of the Renyi entropies for integer $m\ge 2$. The Renyi entropies, per unit time, per particle flux, turn out to be proportional to the total, inclusive cross-section in the initial state, which is free from any IR divergences. The contributions of the small eigenvalues remain subleading, despite the fact that their number grows with the volume in the continuum limit. The entanglement entropy though retains non-analytic, logarithmic behavior with respect to the size of the box, even to all orders in perturbation theory, in the continuum limit. This reveals strong entanglement between the soft and hard particles produced in the scattering process.

It would be interesting to extend our analysis to gravitational and black hole scattering processes, and study the correlation between the soft photons and gravitons with the hard particles produced in the process. Interesting correlations could be uncovered if we examine four-dimensional celestial scattering amplitudes involving boost eigenstates, which are sensitive to both UV and IR physics. The factorization structure of the S-matrix found in \cite{Arkani} could be important in order to quantify the entanglement between the soft and the hard degrees of freedom, and the information carried the unobserved soft particles.

\section*{Acknowledgements} 
We thank P. Betzios for discussions. N.T. thanks the ITCP and the Department of Physics at the University of Crete where parts of this work were carried out for hospitality. 

\bigskip
\appendix
\labelformat{section}{Appendix #1}
\numberwithin{equation}{section}

\section{Notations and conventions}\label{a1}
We work in the Lorenz gauge $\partial_{\mu}A^{\mu}=0$, in which
the electromagnetic gauge field satisfies the wave equation $\partial_\nu\partial^\nu A^{\mu}=0$. We adopt a mostly plus signature metric. In terms of  creation and annihilation operators, the gauge field is given by
\begin{gather}
A^{\mu}(x)=\int\frac{d^3k}{(2\pi)^3}\frac{1}{\sqrt{\omega_{\vec{k}}}}\sum_{r}\left(\epsilon^{\mu}_{r}(\vec{k})a_{r}(\vec{k})e^{ikx}+ \epsilon^{\mu*}_{r}(\vec{k})a^{\dagger}_{r}(\vec{k})e^{-ikx}  \right)  
\end{gather}
with the Greek indices taking values from 0 to 3. We adopt the conventions of \cite{Peskin}.  The polarisation vector $\epsilon^{\mu}_{r}(\vec{k})$ satisfies the following orthonormality relations 
\begin{gather}
\epsilon_{r\mu}(\vec{k})\epsilon^{\mu*}_{s}(\vec{k})=\zeta_{r}\delta_{rs},\ \ \ \sum_{r}\zeta_{r}\epsilon^{\mu}_{r}(\vec{k})\epsilon^{\nu*}_{r}(\vec{k})=\eta^{\mu\nu}
\end{gather}
where $\zeta_0=-1,\ \zeta_1=\zeta_2=\zeta_3=1$.  The commutators of the photon creation and annihilation operators are given by
\begin{gather} 
[a_{r}(\vec{p}),a^{\dagger}_{s}(\vec{q})]=(2\pi)^3\delta^3(\vec{p}-\vec{q})\zeta_{r}\delta_{rs}
\end{gather}
Notice also the Gupta-Bleuler condition on physical states
\begin{gather}
\left[a_{0}(\vec{p}),a_{3}(\vec{p})\right]|\Psi\rangle=0
\end{gather}
Finally the non-trivial anticommutators associated with the electron/positron creation and annihilation operators are, respectively
\begin{gather} 
\{b_{r}(\vec{p}),b^{\dagger}_{s}(\vec{q})\}=\{d_{r}(\vec{p}),d^{\dagger}_{s}(\vec{q})\}=(2\pi)^3\delta^3(\vec{p}-\vec{q})\delta_{rs}
\end{gather}
The indices $r, s, \ldots$ of the spinor creation and annihilation operators take the values 1, 2. We hope they will not be confused with the same symbols used to label the polarisations of the electromagnetic field. 

In the continuum, single particle states are normalized so that they satisfy the Lorentz invariant norm
\begin{equation}
\bra{\vec{q},r}{\vec{p},s}\rangle=2E_{\vec{p}} (2\pi)^3 \delta^3(\vec{p}-\vec{q})\delta_{rs}.
\end{equation} 
On the other hand, box single particle states have discrete momenta
\begin{equation}
\vec{k} = \frac{2\pi}{L}(n_1,\,n_2,\,n_3) 
\end{equation}
and they are unit normalized.

We also list some partial traces over the soft part of the Hilbert space, ${\cal{H}}_S$, associated with dressed ket-bra operators, which are useful in obtaining the reduced density matrix $\rho_H$ -- see \cite{TT} for explicit derivations. We first have
\begin{equation}
\Tr_{{\cal{H}}_S}\left(|\beta\rangle_{d}\langle\beta^\prime|_{d}\right)=|\beta\rangle_{H}\langle\beta^\prime|_{H}~\langle f_{\beta^\prime}|f_\beta\rangle\label{trace1}
\end{equation}
where $\langle f_{\beta^\prime}|f_\beta\rangle$ is given in \ref{braketf}, in the large volume, continuum limit. Taking $|\vec{q}_\gamma|<E_d$ gives
\begin{equation}
\Tr_{{\cal{H}}_S}\left(|\beta\gamma\rangle_{d}\langle\beta^\prime|_{d}\right)=|\beta\rangle_{H}\langle\beta^\prime|_{H}~\langle f_{\beta^\prime}|f_\beta\rangle~\frac{1}{(2V\omega_\gamma)^{1/2}}~\left(f_{\beta^\prime}^*(\vec{q}_\gamma)-f_\beta^*(\vec{q}_\gamma)\right)\cdot \epsilon_r(\vec{q}_\gamma) \label{trace2}
\end{equation}
Finally, when both $|\vec{q}_\gamma|,\,|\vec{q}_{\gamma'}|<E_d$, we obtain
$$
\Tr_{{\cal{H}}_S}\left(|\beta\gamma\rangle_{d}\langle\beta^\prime\gamma^\prime|_{d}\right)~=~|\beta\rangle_{H}\langle \beta^\prime|_{H}~\langle f_{\beta^\prime}|f_\beta\rangle
$$
\begin{equation}
\times~\left\{\delta_{rr^\prime}\delta_{\vec{q}_\gamma\vec{q}_{\gamma'}}+\frac{1}{(2V\omega_{\gamma'})^{1/2}(2V\omega_\gamma)^{1/2}}~\left(f_\beta(\vec{q}_{\gamma'})-f_{\beta^\prime}(\vec{q}_{\gamma'})\right)\cdot \epsilon_{r^\prime}(\vec{q}_{\gamma'})~\left(f_{\beta^\prime}(\vec{q}_\gamma)-f_\beta(\vec{q}_\gamma)\right)\cdot \epsilon_{r}(\vec{q}_\gamma)\right\} \label{trace3}
\end{equation}
Partial traces for the cases in which two or more soft photons are present in the initially undressed states are higher order in the dressing function $f_\beta^{\mu}(\vec{q}_\gamma)$.

\section{Leading order analysis}\label{a2}
\setcounter{equation}{0}
The terms in the expansion of the reduced density matrix \ref{dHard} relevant for the leading order computation of the Renyi and the entanglement entropies are
$$  
\rho_{H}=|\alpha\rangle_{H}\langle\alpha|_{H}+ \bigg( C_{\beta} |\beta\rangle_{H} +\sum_{\omega_\gamma>E}C_{\beta\gamma} |\beta\gamma\rangle_{H} +\dots\bigg) \langle\alpha|_{H} 
$$
$$
+|\alpha\rangle_{H}\bigg( C^{*}_{\beta^{'}} \langle\beta^{'}|_{H} +\sum_{\omega_{\gamma^{'}}>E} C^{*}_{\beta^{'}\gamma^{'}} \langle\beta^{'}\gamma^{'}|_{H} +\dots\bigg) 
$$
$$
+D_{\beta,\beta^{'}} |\beta\rangle_{H}\langle\beta^{'}|_{H}
+\sum_{\omega_{\gamma^{'}},\omega_\gamma>E}D_{\beta\gamma,\beta^{'}\gamma^{'}} |\beta\gamma\rangle_{H}\langle\beta^{'}\gamma^{'}|_{H} 
$$
\begin{equation}
+\sum_{\omega_\gamma>E}D_{\beta\gamma,\beta^{'}} |\beta\gamma\rangle_{H}\langle\beta^{'}|_{H} +\sum_{\omega_{\gamma^{'}}>E}D_{\beta^{'}\gamma^{'},\beta} |\beta\rangle_{H}\langle\beta^{'}\gamma^{'}|_{H} +\dots
\label{reduced}
\end{equation}
where
\begin{equation}  \label{Cbeta}
\frac{C_\beta}{\langle f_\alpha|f_\beta\rangle}= 
\widetilde{T}_{\beta\alpha} + \sum_{\omega_\gamma<E_d}\frac{1}{({2V\omega_\gamma})^{\frac{1}{2}}} 
\widetilde{T}_{\beta\gamma,\alpha} \left( f^{*}_\alpha(\vec{q_\gamma})- f^{*}_\beta(\vec{q_\gamma}) \right)\cdot\epsilon(\gamma)+\dots
\end{equation}
\begin{equation}  \label{Cbetagamma}
C_{\beta\gamma}=\langle f_\alpha|f_\beta\rangle 
\widetilde{T}_{\beta\gamma,\alpha} +\dots
\end{equation}
$$  
\frac{D_{\beta,\beta^{'}}}{\langle f_{\beta^{'}}|f_\beta\rangle}= \widetilde{T}_{\beta\alpha}\widetilde{T}^{*}_{\beta^{'}\alpha}+ \sum_{\omega_\gamma<E_d}\frac{1}{({2V\omega_\gamma})^{\frac{1}{2}}}\widetilde{T}_{\beta\gamma,\alpha}\widetilde{T}^{*}_{\beta^{'}\alpha} \left( f^{*}_{\beta^{'}}(\vec{q_\gamma})- f^{*}_\beta(\vec{q_\gamma}) \right)\cdot\epsilon(\gamma)
$$
\begin{equation}
+\sum_{\omega_\gamma^{'}<E_d}\frac{1}{({2V\omega_\gamma})^{\frac{1}{2}}}\widetilde{T}^{*}_{\beta^{'}\gamma^{'},\alpha}\widetilde{T}_{\beta\alpha} \left( f_\beta(\vec{q}_{\gamma^{'}})- f_{\beta^{'}}(\vec{q}_{\gamma^{'}}) \right)\cdot\epsilon^{*}(\gamma^{'}) +\sum_{\omega_\gamma<E} \widetilde{T}_{\beta\gamma,\alpha}\widetilde{T}^{*}_{\beta^{'}\gamma,\alpha}+\dots
\end{equation}
\begin{equation} 
D_{\beta\gamma,\beta^{'}}=\langle f_{\beta^{'}}|f_\beta\rangle \widetilde{T}_{\beta\gamma,\alpha}\widetilde{T}^{*}_{\beta^{'}\alpha} +\dots
\end{equation}
\begin{equation}  
D_{\beta\gamma,\beta^{'}\gamma^{'}}=\langle f_{\beta^{'}}|f_\beta\rangle  \widetilde{T}_{\beta\gamma,\alpha}\widetilde{T}^{*}_{\beta^{'}\gamma^{'},\alpha} + \dots 
\end{equation}

Notice that the leading contributions in $C_\beta$ are of order $e^{\cal{N}}$, in $C_{\beta\gamma}$ of order $e^{{\cal{N}}+1}$, in  $D_{\beta,\beta^{'}}$ of order $e^{2{\cal{N}}}$, in  $D_{\beta\gamma,\beta^{'}}$ of order $e^{2{\cal{N}}+1}$ and in $D_{\beta\gamma,\beta^{'}\gamma^{'}}$ of order $e^{2{\cal{N}}+2}$.  Also in the expression for $C_\beta$, the second term and terms in the ellipses vanish when $\beta=\alpha$ \cite{TT}. So $C_\alpha=\widetilde{T}_{\alpha\alpha}$ to all orders. 

Let us verify that $\Tr\rho_{H}=\Tr\rho=1$ using unitarity, \ref{unitarity}. We have
$$
\Tr\rho_H-1=C_\alpha+C^{*}_\alpha+\sum_{\beta}\bigg( D_{\beta,\beta}+\sum_{\omega_\gamma>E}D_{\beta\gamma,\beta\gamma}\bigg)+\dots
$$
\begin{equation}
=\widetilde{T}_{\alpha\alpha}+\widetilde{T}^{*}_{\alpha\alpha}+\sum_{\beta}\widetilde{T}_{\beta\alpha}\widetilde{T}^{*}_{\beta\alpha}+\sum_{\beta\gamma}\widetilde{T}_{\beta\gamma,\alpha}\widetilde{T}^{*}_{\beta\gamma,\alpha}+\dots=_d\langle\alpha|i(T-T^{\dagger})+T^{\dagger}T|\alpha\rangle_{d}=0
\end{equation}
Therefore, $C_\alpha+C^{*}_\alpha=\widetilde{T}_{\alpha\alpha}+\widetilde{T}^{*}_{\alpha\alpha}$ is of order $e^{4}$ by unitarity.

\subsection{Traces to leading order}
To facilitate the perturbative analysis, we set
\begin{equation}  
\rho_H=\rho_0+\varepsilon,\ \ \ \rho_0=|\alpha\rangle_{H}\langle\alpha|_{H}
\end{equation}
Since $\varepsilon$ is of order $e^2$,  we need to expand $(\rho_{H})^{m}$ to cubic order in order to obtain the leading perturbative contributions to the Renyi entropies. The fact that $(\rho_0)^2=\rho_0$ and the cyclic property of the trace limit the number of structures we need to consider. As in the case of two-electron scattering, only linear and quadratic structures contribute to leading order ($e^6$). The cubic structures in $\varepsilon$ turn out to vanish (to order $e^6$). 

To linear order in $\varepsilon$, it suffices to consider the term $\varepsilon\rho_0$ whose trace is
\begin{equation}  
\Tr\varepsilon\rho_{0}=C_\alpha+C^{*}_\alpha+D_{\alpha,\alpha}=\widetilde{T}_{\alpha\alpha}+\widetilde{T}^{*}_{\alpha\alpha}+\widetilde{T}_{\alpha\alpha}\widetilde{T}^{*}_{\alpha\alpha}+\sum_{\omega_\gamma<E}\widetilde{T}_{\beta\gamma,\alpha}\widetilde{T}^{*}_{\beta\gamma,\alpha}+\dots
\end{equation}
The ellipses include terms of higher order than $e^{6}$ and do not contribute to the entanglement entropy at leading order.

The quadratic terms include $\varepsilon^2,\varepsilon^2\rho_{0},\varepsilon\rho_{0}\varepsilon\rho_{0}$. The trace of the first term is
$$
\Tr\varepsilon^2=C^2_\alpha+C^{*2}_\alpha+\sum_{\beta}2D_{\alpha,\beta}C_\beta+2D_{\beta,\alpha}C^{*}_\beta+2C_{\beta}C^{*}_\beta+2\sum_{\omega_\gamma>E}C_{\beta\gamma,\alpha}C^{*}_{\beta\gamma,\alpha}+\dots
$$
$$
=\widetilde{T}^2_{\alpha\alpha}+\widetilde{T}^{*2}_{\alpha\alpha}+2\sum_{\beta}|\langle f_\beta|f_\alpha\rangle|^2\left(1+\widetilde{T}_{\alpha\alpha}+\widetilde{T}^{*}_{\alpha\alpha}\right)\widetilde{T}_{\beta\alpha}\widetilde{T}^{*}_{\beta\alpha}
$$
$$
+2\sum_{\beta}\sum_{\omega_\gamma<E_d}\frac{|\langle f_\beta|f_\alpha\rangle|^2}{(2V\omega_\gamma)^{1/2}}
\times\left(\widetilde{T}_{\beta\alpha}\widetilde{T}^{*}_{\beta\gamma,\alpha}\left(f_\alpha(\vec{q}_{\gamma})- f_\beta(\vec{q}_{\gamma})\right)\cdot\epsilon^{*}(\gamma)+ \widetilde{T}^{*}_{\beta\alpha}\widetilde{T}_{\beta\gamma,\alpha}\left(f^{*}_\alpha(\vec{q}_{\gamma})- f^{*}_\beta(\vec{q}_{\gamma})\right)\cdot\epsilon(\gamma) \right)
$$
\begin{equation}
+2\sum_{\beta}\sum_{\omega_\gamma>E}|\langle f_\beta|f_\alpha\rangle|^2\widetilde{T}_{\beta\gamma,\alpha}\widetilde{T}^{*}_{\beta\gamma,\alpha}+\dots
\label{epsilon2}
\end{equation}
The third line is absent for the Fock basis computation.

For the second quadratic trace we obtain
$$
\Tr (\varepsilon^2\rho_0) = C_\alpha^2 + C_\alpha^{*\, 2} + C_{\alpha}C_{\alpha}^*+ D_{\alpha,\alpha} (C_\alpha + C_{\alpha}^*)
$$
$$
+ \sum_\beta \bigg(D_{\alpha,\beta}C_\beta +
D_{\beta,\alpha}C_{\beta}^*+ C_\beta C_\beta^*~+~\sum_{\omega_\gamma>E}C_{\beta\gamma}C_{\beta\gamma}^*\bigg) + \dots
$$
$$
= \widetilde{T}_{\alpha\alpha}^2 + \widetilde{T}_{\alpha\alpha}^{*\, 2}+\left(1+\widetilde{T}_{\alpha\alpha}+\widetilde{T}_{\alpha\alpha}^*\right) \widetilde{T}_{\alpha\alpha}\widetilde{T}_{\alpha\alpha}^*+\sum_{\beta} |\bra{f_\beta}\ket{f_\alpha}|^2\left(1+\widetilde{T}_{\alpha\alpha}+\widetilde{T}_{\alpha\alpha}^*\right)\widetilde{T}_{\beta\alpha}\widetilde{T}_{\beta\alpha}^*
$$
$$
+\sum_{\beta}\sum_{\omega_\gamma<{E_d}}\frac{|\bra{f_\beta}\ket{f_\alpha}|^2}{(2V\omega_\gamma)^{1/2}} \left[ \widetilde{T}_{\beta\alpha}\widetilde{T}_{\beta\gamma,\,\alpha}^*\left(f_\alpha(\vec{q}_\gamma)-f_\beta(\vec{q}_\gamma)\right)\cdot \epsilon^*(\gamma)+\widetilde{T}_{\beta\alpha}^*\widetilde{T}_{\beta\gamma,\,\alpha}\left(f_\alpha^*(\vec{q}_\gamma)-f_\beta^*(\vec{q}_\gamma)\right)\cdot \epsilon(\gamma)\right]
$$
\begin{equation}
+\sum_\beta\sum_{\omega_\gamma>E}|\bra{f_\beta}\ket{f_\alpha}|^2\widetilde{T}_{\beta\gamma,\,\alpha}\widetilde{T}_{\beta\gamma,\,\alpha}^* +\dots
\end{equation}

Notice the appearance of off diagonal elements in the perturbative expansions for both $\Tr \varepsilon^2$ and $\Tr \varepsilon^2\rho_0$. At any finite order in perturbation theory, the off diagonal elements are non-zero and contain IR logarithmic divergences in $\lambda$. Both of these traces contribute to the Renyi and the entanglement entropies to leading order ($e^6$).

Finally we have
$$
\Tr\varepsilon\rho_{0}\varepsilon\rho_{0}=(C_\alpha+C^{*}_\alpha)(C_\alpha+C^{*}_\alpha+2D_{\alpha,\alpha})+\dots 
$$
\begin{equation}
=(\widetilde{T}_{\alpha\alpha}+\widetilde{T}^{*}_{\alpha\alpha})(\widetilde{T}_{\alpha\alpha}+\widetilde{T}^{*}_{\alpha\alpha}+2\widetilde{T}_{\alpha\alpha}\widetilde{T}^{*}_{\alpha\alpha})+\dots
\end{equation}
which does not contribute to order $e^{6}$ by unitarity.

The cubic terms are $\varepsilon^3$, $\varepsilon^3\rho_0$, $\varepsilon^2\rho_{0}\varepsilon\rho_0$, $\varepsilon\rho_{0}\varepsilon\rho_{0}\varepsilon\rho_{0}$. However these terms do not contribute to leading order. Firstly
$$
\Tr\varepsilon^3=C^{3}_\alpha+C^{*3}_\alpha+ 3(C_\alpha+C^{*}_\alpha)\sum_{\beta}C_{\beta}C^{*}_\beta+\dots
$$
\begin{equation}
=\widetilde{T}^{3}_{\alpha\alpha}+\widetilde{T}^{*3}_{\alpha\alpha}+3(\widetilde{T}_{\alpha\alpha}+\widetilde{T}^{*}_{\alpha\alpha})\sum_{\beta}C_\beta C^{*}_\beta +\dots
\end{equation}
which does not contribute to order $e^{6}$ by unitarity. 

For the next terms we have
\begin{equation}  
\Tr(\varepsilon^3\rho_0)=C^{3}_\alpha+C^{*3}_\alpha+ (C_\alpha+C^{*}_\alpha)\Big(C_{\alpha}C^{*}_\alpha+2\sum_{\beta}C_{\beta}C^{*}_\beta\Big)+\dots
\end{equation}
\begin{equation}  
\Tr(\varepsilon^2\rho_0\varepsilon\rho_0)=(C_{\alpha}+C^{*}_\alpha) \Big(C^{2}_\alpha+C^{*2}_\alpha+C_{\alpha}C^{*}_\alpha+\sum_{\beta}C_{\beta}C^{*}_\beta\Big)+\dots
\end{equation}
\begin{equation}  
\Tr(\varepsilon\rho_0\varepsilon\rho_0\varepsilon\rho_0)=(C_{\alpha}+C^{*}_\alpha)^3+\dots
\end{equation}
which do not contribute to the leading entanglement entropy.

To order $e^{6}$, only $\Tr(\varepsilon\rho_0)$, $\Tr(\varepsilon^2)$ and $\Tr(\varepsilon^2\rho_0)$ contribute.

To proceed, we use the perturbative expansion for the overlap of the coherent states \ref{braketf}. Recall that we first expand to a given order in perturbation theory, keeping the volume of the box and the cutoff $\lambda$ finite. The continuum limit (where the volume of the box is taken to infinity and the cutoff $\lambda$ to zero) is taken in the end. If the logarithmic IR divergences do not cancel, the perturbative expansion breaks down in the continuum limit. We will also need the relations between dressed and undressed amplitudes \ref{du1} and \ref{du2}. We will use these relations to express the traces in terms of undressed amplitudes, which we can calculate more easily via the Feynman diagrams. 

For the first trace we have
\begin{equation}  \label{epsilonrho}
\Tr\varepsilon\rho_{0}=T_{\alpha\alpha}+T^{*}_{\alpha\alpha}+T_{\alpha\alpha}T^{*}_{\alpha\alpha}+\bigg(\sum_{\omega_\gamma<E_d}\frac{1}{2V\omega_\gamma}F_{\alpha\alpha}(\gamma)F^{*}_{\alpha\alpha}+ \sum_{E_d<\omega_\gamma<E}T_{\alpha\gamma,\alpha}T^{*}_{\alpha\gamma,\alpha}\bigg)
\end{equation}
The second term in parentheses is zero due to energy conservation. In the continuum limit, the first term in parentheses yields 
\begin{equation}  \label{eq:2.41}
\int^{E_d}_{\lambda}\frac{d^3\vec{q}}{(2\pi)^3 2\omega_{\vec{q}}}\sum_{r}|F_{\alpha\alpha}(\vec{q},\epsilon_{r}(\vec{q}))|^2
\end{equation}
up to $\lambda$ independent multiplicative factors.
Since $F_{\alpha\alpha}(\vec{q},\epsilon_{r}(\vec{q}))$ is finite and nonsingular in the limits $|\vec{q}|,\lambda\rightarrow0$, the contribution of this term is at most of order $E^2_d$, and can be neglected relative to other contributions. 

For the quadratic traces we obtain
$$  
\Tr\varepsilon^2=T^2_{\alpha\alpha}+T^{*2}_{\alpha\alpha}+2\sum_{\beta}\bigg(T_{\beta\alpha}T^{*}_{\beta\alpha}+\sum_{\omega_\gamma>E}T_{\beta\gamma,\alpha}T^{*}_{\beta\gamma,\alpha}\bigg)
$$
\begin{equation}
+2\sum_{\beta}\sum_{\omega_\gamma<E_d}\frac{1}{2V\omega_\gamma}
\left[T_{\beta\alpha}F^{*}_{\beta\alpha}(\gamma)\left(f_\alpha(\vec{q}_{\gamma})- f_\beta(\vec{q}_{\gamma})\right)\cdot\epsilon^{*}(\gamma)+ T^{*}_{\beta\alpha}F_{\beta\alpha}\left(f^{*}_\alpha(\vec{q}_{\gamma})- f^{*}_\beta(\vec{q}_{\gamma})\cdot\epsilon(\gamma)\right)\right]
\label{epsilon2}
\end{equation}

$$   
\Tr(\varepsilon^2\rho_0)=T^2_{\alpha\alpha}+T^{*2}_{\alpha\alpha}+T_{\alpha\alpha}T^{*}_{\alpha\alpha}+2\sum_{\beta}\bigg(T_{\beta\alpha}T^{*}_{\beta\alpha}+\sum_{\omega_\gamma>E}T_{\beta\gamma,\alpha}T^{*}_{\beta\gamma,\alpha}\bigg)
$$
\begin{equation}
+2\sum_{\beta}\sum_{\omega_\gamma<E_d}\frac{1}{2V\omega_\gamma}\left[T_{\beta\alpha}F^{*}_{\beta\alpha}(\gamma)\left(f_\alpha(\vec{q}_{\gamma})- f_\beta(\vec{q}_{\gamma})\right)\cdot\epsilon^{*}(\gamma)+ T^{*}_{\beta\alpha}F_{\beta\alpha}\left(f^{*}_\alpha(\vec{q}_{\gamma})- f^{*}_\beta(\vec{q}_{\gamma}).\epsilon(\gamma)\right)\right]
\label{epsilon2rho}
\end{equation}

Since $F_{\beta\alpha}$ is of order $e^{3}$ and $f_\alpha$ is of order $e$, $T_{\beta\alpha}$ must be calculated at tree level. So it does not lead to any IR divergences in the limit $\lambda\rightarrow0$. 
In the continuum limit, the last lines of expressions \ref{epsilon2} and \ref{epsilon2rho} are given by the following integral (up to smooth factors in the limit $\lambda \to 0$ and volume factors)
\begin{equation}   
\int^{E_d}_{\lambda}\frac{d^3\vec{q}}{(2\pi)^3 2\omega_{\vec{q}}}\sum_{r}F^{*}_{\beta\alpha}(\vec{q}_\gamma,\epsilon_{r}(\vec{q}_\gamma))\sum_{s\in\{\alpha,\beta\}}\frac{e_{s}\eta_{s}p_{s}\cdot\epsilon^{*}_{r}(\vec{q})}{p_{s}\cdot q}+h.c.
\end{equation}
The integral is not divergent in the limit $|\vec{q}|\rightarrow0$. In fact it is of order $E_d$ making a negligible contribution to the Renyi and the entanglement entropies.

Let us calculate the quantity $\Tr(\rho_H)^2$ to order $e^{6}$. 
$$  
\Tr(\rho_H)^2=\Tr(\rho_0)^2+2\Tr\varepsilon\rho_0+\Tr\varepsilon^2
$$
\begin{equation}
=1+2(T_{\alpha\alpha}+T^{*}_{\alpha\alpha})+(T_{\alpha\alpha}+T^{*}_{\alpha\alpha})^2+2\sum_{\beta}\bigg(T_{\beta\alpha}T^{*}_{\beta\alpha}+\sum_{\omega_{\gamma}>E}T_{\beta\gamma,\alpha}T^{*}_{\beta\gamma,\alpha}\bigg)
\label{rhoH2}
\end{equation}

Using unitarity, \ref{rhoH2} is simplified to
\begin{equation}   
\Tr(\rho_H)^2=1-2\Delta
\end{equation}
where
\begin{equation}   
\Delta=\sum_{\beta}\sum_{\omega_\gamma<E}T_{\beta\gamma,\alpha}T^{*}_{\beta\gamma,\alpha}
\end{equation}
is of order $e^{6}$, depending on the amplitude for single real photon emission in the energy range $\lambda<\omega_\gamma<E$. 

Next we calculate $\Tr(\rho_H)^m$ with $m\geq3$ to order $e^{6}$. We obtain
\begin{equation}   \label{rhoHm}
\Tr(\rho_H)^m=1+m\Tr\varepsilon\rho_0+m\Tr\varepsilon^2\rho_0
\end{equation}
which simplifies further to
\begin{equation}   
\Tr(\rho_H)^m=1-m\Delta
\end{equation}
To leading order, the Renyi entropies for integer $m\geq 2$ are given by
\begin{equation}
S_{m}=\frac{1}{1-m}\ln\Tr(\rho_H)^{m}=\frac{m}{1-m}\Delta
\end{equation}

\subsection{The large eigenvalue to leading order}
The large eigenvalue of $\rho_H$ is given by 
$$
\lambda_{\Phi}=\bra{\Phi}\ket{\Phi}=1+C_{\alpha}+C^*_{\alpha}+\sum_{\beta}C_{\beta}C^*_{\beta}+\sum_{\beta}\sum_{\omega_\gamma>E}C_{\beta\gamma}C^*_{\beta\gamma}+\dots
$$
$$
=1+ \widetilde{T}_{\alpha\alpha}+\widetilde{T}^*_{\alpha\alpha}+\sum_{\beta}|\bra{f_\alpha}\ket{f_\beta}|^2\widetilde{T}_{\beta\alpha}\widetilde{T}^*_{\beta\alpha} + \sum_{\beta}\sum_{\omega_\gamma>E}|\bra{f_\alpha}\ket{f_\beta}|^2\widetilde{T}_{\beta\gamma,\alpha}\widetilde{T}^*_{\beta\gamma,\alpha}
$$
\begin{equation}
+\bigg(\sum_{\beta}\sum_{\omega_\gamma<E_d}\frac{|\bra{f_\alpha}\ket{f_\beta}|^2}{(2V\omega_\gamma)^{1/2}} \widetilde{T}^*_{\beta\alpha} \widetilde{T}_{\beta\gamma,\alpha} \Big( f^{*}_\alpha(\vec{q_\gamma})- f^{*}_\beta(\vec{q_\gamma}) \Big)\cdot\epsilon(\gamma)~+~h.c.\bigg)~+~{\cal{O}}(e^8)
\end{equation}
As shown in \cite{TT} (and also discussions in the previous subsection), the terms in the last line give suppressed contributions, of order the characteristic energy of the photons in the clouds $E_d$, and can be dropped. Using unitarity and the relations between dressed and undressed amplitudes (\ref{du1} and \ref{du2}), $\lambda_{\Phi}$ can be written as
\begin{equation}
\lambda_{\Phi}=1+ T_{\alpha\alpha}+T^*_{\alpha\alpha}+\sum_{\beta}T_{\beta\alpha}T^*_{\beta\alpha} + \sum_{\beta}\sum_{\omega_\gamma>E}T_{\beta\gamma,\alpha}T^*_{\beta\gamma,\alpha}~+~{\cal{O}}(e^8)=1-\Delta~+~{\cal{O}}(e^8)
\end{equation}

The other nonvanishing eigenvalues coincide with the nonvanishing eigenvalues of the matrix $G$, and each is of order $e^6$, at least. Their sum is equal to $\Delta$. Let us verify this by computing explicitly the trace of $G$ to order $e^6$:
\begin{equation} \label{G1} 
\Tr \,G = \sum_{\beta} \Big(1-\bra{f_\alpha}{f_\beta}\rangle\bra{f_\alpha}{f_\beta}{\rangle}^*\Big)\widetilde{T}_{\beta\alpha}\widetilde{T}^*_{\beta\alpha} + \sum_{\beta}\sum_{\omega_\gamma<E}\widetilde{T}_{\beta\gamma,\alpha}\widetilde{T}^*_{\beta\gamma,\alpha}
\end{equation}  
Using unitarity, this can be written as
\begin{equation} \label{G2}
\Tr\, G = -T_{\alpha\alpha}-T^*_{\alpha\alpha} -\sum_{\beta}\sum_{\omega_\gamma>E}T_{\beta\gamma,\alpha}T^*_{\beta\gamma,\alpha}-\sum_{\beta}T_{\beta\alpha}T^*_{\beta\alpha}=\Delta 
\end{equation}


\section{The leading entanglement entropy in the continuum limit}\label{a3}
\setcounter{equation}{0}
In the generic case, non-trivial disconnected diagrams in which only two particles interact with each other contribute to the singular part of the leading entanglement entropy, \ref{SENTCONT}, in the continuum, large volume limit. Each disconnected line associated with a noninteracting incoming (outgoing) particle $i$ ($j$), with momentum $\vec{p}_{i,in}$ ($\vec{p}_{j,out}$), contributes to the Feynman diagram with a factor $(2\pi)^3\,2E_{i,in}\,\delta^3(\vec{p}_{i,in}-\vec{p}_{j,out})$.  We consider the case where the momenta of the initial particles differ from each other.

\begin{figure}
\begin{center}
\includegraphics [scale=1.00]{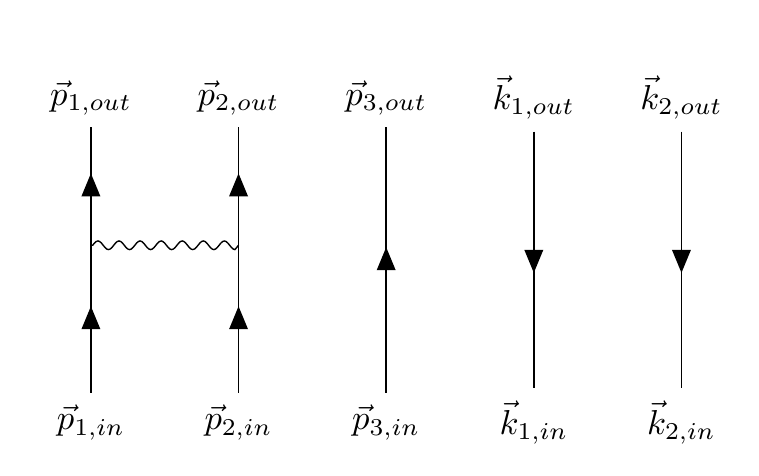}
\caption{\it \footnotesize A disconnected Feynman diagram contributing to $I_1$ for an initial state of 3 electrons and 2 positrons. Additional contributions can be obtained by permuting the outgoing momenta among the external lines associated with the outgoing particles. Diagrams in which the first and third electron, and also the second and third electron must be also added.}
\end{center}
\end{figure}

The amplitude $i{\cal{M}}(\alpha \to me^-+ne^+)$ can be written as a sum of the following three terms. The first term arises from Feynman diagrams in which only two electrons interact, while the rest of the initial particles do not:
\begin{gather}
I_1=\sum_{i=1}^{m}\sum_{j< i}^{m}\sum_{\sigma,\,\sigma'}\,{\rm{sign}}(\sigma)\,{\rm{sign}}(\sigma')\,\prod_{l\neq i,j}^{m}\bigg((2\pi)^32E(\vec{p}_{l,in})\delta^3(\vec{p}_{l,in}-\vec{p}_{\sigma(l),out})\bigg)\nonumber\\
\times\,\prod_{l'=1}^{n}\bigg((2\pi)^32E(\vec{k}_{l',in})\delta^3(\vec{k}_{l',in}-\vec{k}_{\sigma'(l'),out})\bigg)\, i\mathcal{M}_{1t}\Big(\vec{p}_{i,in},\vec{p}_{j,in};\vec{p}_{\sigma(i),out},\vec{p}_{\sigma(j),out}\Big)
\end{gather}
We have denoted the momenta of the electrons by $\vec{p}_l$ and of the positrons by $\vec{k}_{l'}$. For brevity, we have omitted the inclusion of polarizations indices, since we average over these in the end. In figure 2 we show a disconnected diagram contributing to $I_1$ to leading order, for an initial state of 3 electrons and 2 positrons. Also, $i\mathcal{M}_1$ is the scattering amplitude for the process $e^-e^-\to e^-e^-$, calculated at tree level, and $i\mathcal{M}_{1t}$ is the $t$-channel amplitude. The permutation of $m$ ($n$) objects is denoted by  $\sigma$ ($\sigma'$).

The second term arises from Feynman diagrams in which only two positrons interact:
\begin{gather}
I_2=\sum_{i=1}^{n}\sum_{j< i}^{n}\sum_{\sigma,\,\sigma' }\,{\rm{sign}}(\sigma)\,{\rm{sign}}(\sigma')\,\prod_{l=1}^{m}\bigg((2\pi)^32E(\vec{p}_{l,in})\delta^3(\vec{p}_{l,in}-\vec{p}_{\sigma(l),out})\bigg)\nonumber\\
\times\,\prod_{l'\neq i,j}^{n}\bigg((2\pi)^32E(\vec{k}_{l',in})\delta^3(\vec{k}_{l',in}-\vec{k}_{\sigma'(l'),out})\bigg)\, i\mathcal{M}_{2t}\Big(\vec{k}_{i,in},\vec{k}_{j,in};\vec{k}_{\sigma'(i),out},\vec{k}_{\sigma'(j),out}\Big)
\end{gather}
where $i\mathcal{M}_2$ is the ampltude for the process $e^+e^+\to e^+e^+$ and $i\mathcal{M}_{2t}$ is the $t$-channel amplitude. See figure 3.

\begin{figure}
\begin{center}
\includegraphics [scale=1.00]{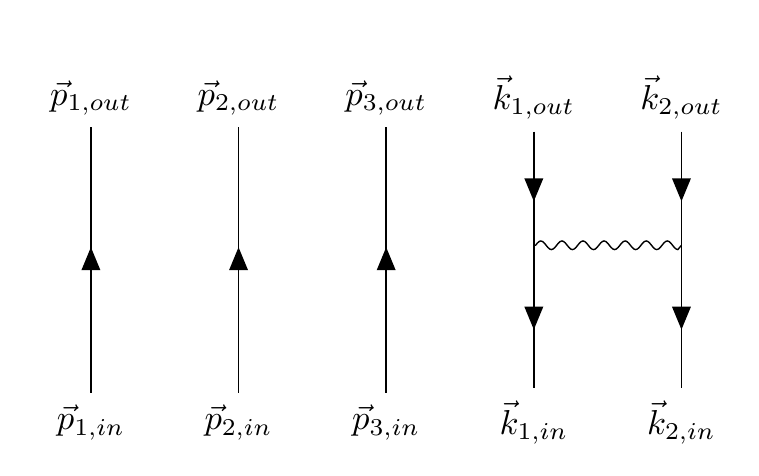}
\caption{\it \footnotesize A disconnected Feynman diagram contributing to $I_2$ for an initial state of 3 electrons and 2 positrons.}
\end{center}
\end{figure}

Finally, when one electron and one positron interact, we get the following contribution
\begin{gather}
I_3=\sum_{i=1}^{m}\sum_{j=1}^{n}\sum_{\sigma,\,\sigma' }\,{\rm{sign}}(\sigma)\,{\rm{sign}}(\sigma')\,\prod_{l\neq i}^{m}\bigg((2\pi)^32E(\vec{p}_{l,in})\delta^3(\vec{p}_{l,in}-\vec{p}_{\sigma(l),out})\bigg)\nonumber\\
\times\,\prod_{l'\neq j}^{n}\bigg((2\pi)^32E(\vec{k}_{l',in})\delta^3(\vec{k}_{l',in}-\vec{k}_{\sigma'(l'),out})\bigg)\, i\mathcal{M}_{3}\Big(\vec{p}_{i,in},\vec{k}_{j,in};\vec{p}_{\sigma(i),out},\vec{k}_{\sigma'(j),out}\Big)
\end{gather}
where $i\mathcal{M}_3$ is the total amplitude for the process $e^-e^+\to e^-e^+$, calculated at tree level. See figure 4 for the t-channel and s-channel diagrams. 

To leading order, the amplitude $i{\cal{M}}(\alpha \to (m-1)e^-+(n-1)e^++2\gamma)$ takes the form
\begin{gather}
\sum_{i=1}^{m}\sum_{j=1}^{n}\sum_{\sigma,\,\sigma' }\,{\rm{sign}}(\sigma)\,{\rm{sign}}(\sigma')\,
\prod_{l\neq i}^{m}\bigg((2\pi)^32E(\vec{p}_{l,in})\delta^3(\vec{p}_{l,in}-\vec{p}_{\sigma(l),out})\bigg)\nonumber\\
\times\,\prod_{l'\neq j}^{n}\bigg((2\pi)^32E(\vec{k}_{l',in})\delta^3(\vec{k}_{l,in}-\vec{k}_{\sigma'(l'),out})\bigg)
\, i\mathcal{M}_4\Big(\vec{p}_{i,in},\vec{k}_{j,in};\vec{q}_{1,out},\vec{q}_{2,out}\Big)
\end{gather}
where $i\mathcal{M}_4$ is the tree-level amplitude for the process  $e^-e^+\to 2\gamma$. The momenta of the photons are denoted by $\vec{q}_1$ and $\vec{q}_2$. Relevant diagrams are shown in figure 5.

\begin{figure}
\begin{center}
\includegraphics [scale=0.85]{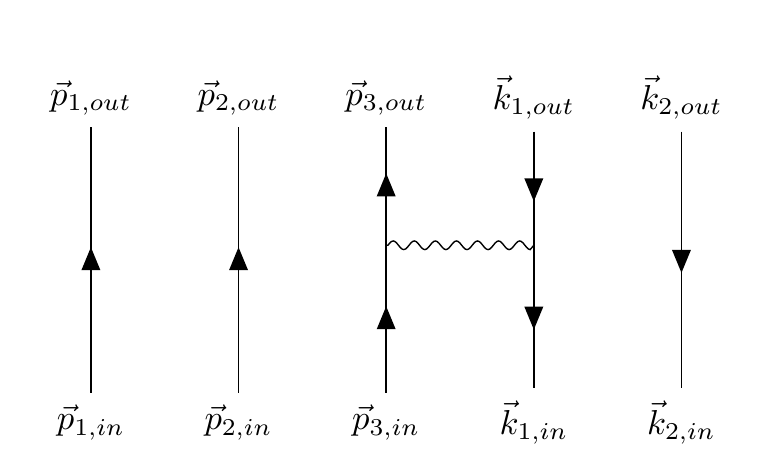}
\,\,\,\,\,\,\,\,\,\,
\includegraphics [scale=0.65]{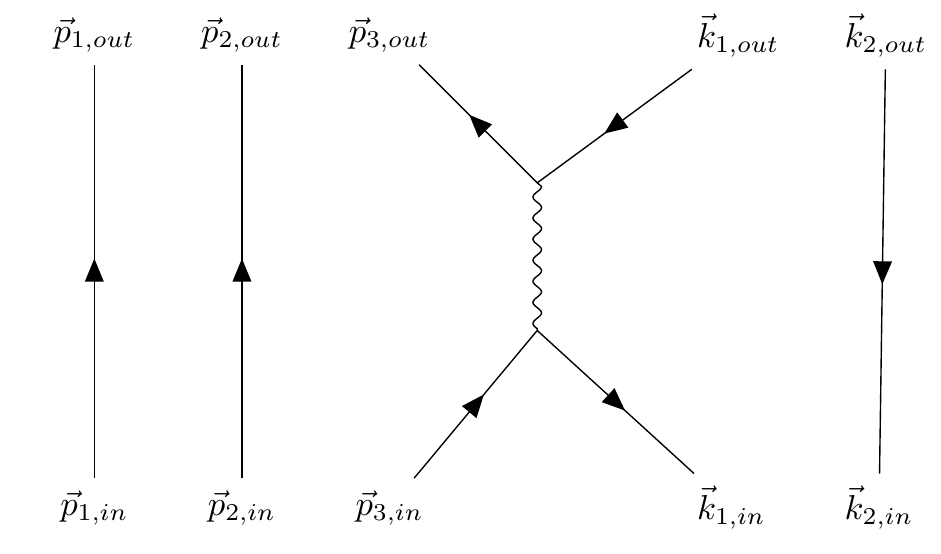}
\caption{\it \footnotesize Disconnected Feynman diagrams contributing to $I_3$. On the left is the t-channel contribution and on the right the s-channel contribution.}
\end{center}
\end{figure}

To calculate the entanglement entropy, we need to determine the square of the absolute value of the amplitudes $i{\cal{M}}(\alpha \to me^-+ne^+)$ and $i{\cal{M}}(\alpha\to (m-1)e^-+(n-1)e^++2\gamma)$, and integrate over the momenta of the final particles (\ref{SENTCONT}). Each amplitude is obtained as a sum of Feynman diagrams in which only a pair of particles interact, as we explained before. Taking the absolute value squared, a number of terms appear, each containing a product of $2(m+n-2)$ $\delta$-functions (in three-dimensional momentum space). There is also an overall $4$-momentum $\delta$-function squared, imposing energy-momentum conservation -- see \ref{SENTCONT}. In most cross-terms that are not perfect squares, the $\delta$-functions are incompatible with each other. So, upon integrating over the momenta of the noninteracting particles, such cross terms give vanishing contributions.   

The dominant contributions come from perfect squares (of the form $\alpha\alpha^*$). For such terms, the sign of the diagram does not contribute.
For such terms, the products of $\delta$-functions from each factor are identical. If we integrate over the momenta of the $n+m-2=N-2$ noninterracting particles, we get a factor $[(2\pi)^3\delta^3(0)]^{N-2}=V^{N-2}$. 
There are also cross terms with nonvanishing contributions, but these are suppressed in the continuum, large volume limit, since they are proportional to a smaller power of the volume $V$. We will drop such subleading terms. Notice however that the precise volume dependence is non-trivial. A detailed example with three charged particles in the initial state is presented in the following subsection. 

\begin{figure}
\begin{center}
\includegraphics [scale=0.70]{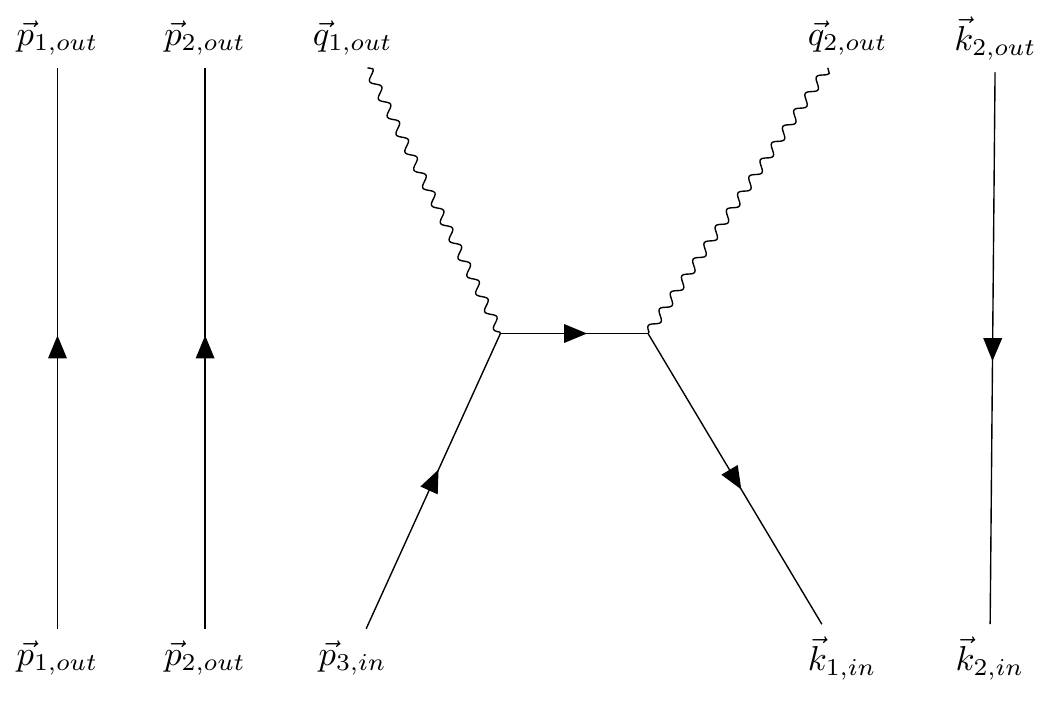}
\caption{\it \footnotesize A disconnected Feynman diagram contributing to $i{\cal{M}}(3e^-+2e^+ \to 2e^-+e^++2\gamma)$. The u-channel diagram is obtained by exchanging the outgoing photon momenta between the outgoing photon lines.}
\end{center}
\end{figure}

For a given pair of interacting incoming electrons, $i$ and $j$, there are $m!n!/2$ dominant terms, each proportional to $\abs{i\mathcal{M}_1(\vec{p}_i,\vec{p}_j;\vec{p}_{i'},\vec{p}_{j'})}^2$, where we denote the corresponding, outgoing interacting particles by $i'$, $j'$. Likewise, we get $m!n!/2$ dominant terms, proportional to $\abs{i\mathcal{M}_2(\vec{p}_i,\vec{p}_j;\vec{p}_{i'},\vec{p}_{j'})}^2$, for a pair of interacting positrons. When the $i$-th electron scatters against the $j$-th positron, we get $m!n!$ dominant terms, each proportional to $\abs{i\mathcal{M}_3(\vec{p}_i,\vec{p}_j;\vec{p}_{i'},\vec{p}_{j'})}^2$. Finally, there are $(m-1)!(n-1)!$ dominant terms produced, proportional to $\abs{i\mathcal{M}_4(\vec{p}_i,\vec{p}_j;\vec{p}_{i'},\vec{p}_{j'})}^2$, when the $i$-th electron and the $j$-th positron annihilate to produce two outgoing photons. In this latter case $i'$, $j'$ denote the two outgoing photons.

There are two more integrations involved in \ref{SENTCONT}, over the momenta of the interacting outgoing particles. Integrating over $\vec{p}_{j'}$, imposes momentum conservation and gives rise to an additional volume factor. We can always choose to work in the center of mass frame for the two interacting incoming particles $i$ and $j$ ($\vec{p}_i=-\vec{p}_j$). The center of mass energy of these particles is denoted by $E_{ij}=2E_i$ and the relative velocity by $v_{ij}=4|\vec{p}_i|/E_{ij}$. We also denote the scattering angle by $\theta$. Integrating over the magnitude of $\vec{p}_{i'}$ results in a factor $(2\pi)\delta(2E_i-2E_i)=T$, equal to the time scale of the experiment, and a factor of the relative velocity $v_{ij}$.  

\subsection{Three-particle scattering}
In this appendix we illustrate the continuum, large volume analysis of \ref{cl} for a particular case concerning a three-electron initial sate. We take the momenta of the incoming particles to be different. The leading perturbative entanglement entropy takes the form
\begin{equation}\label{sent3}
S_{ent,sing}=-\frac{2\ln{e^6}}{3!V^3} \ln\left(\frac{E_d}{\lambda}\right) \bigg(\prod_{f=1}^3\int\frac{d^3\vec{p}_f}{(2\pi)^32E_f}\bigg)\bigg(\prod_{i=1}^3\frac{1}{2E_i}\bigg) ~\Big|i\mathcal{M}(\alpha \to 3e^-)\Big|^2\mathcal{B}_{\beta\alpha}
\Big[(2\pi)^4\delta^4(\sum_fp_f-\sum_ip_i)\Big]^2
\end{equation}
 
The leading contributions to the amplitude $i\mathcal{M}(\alpha \to 3e^-)$ (arising from disconnected tree-level Feynman diagrams in which only a pair of two electrons are interacting) are given by
$$
\Big(i\mathcal{M}_1(p_{1},p_{2};p'_{1},p'_{2})(2\pi)^32E_{3}\delta^3(\vec{p}_{3}-\vec{p}'_{3})+i\mathcal{M}_1(p_{1},p_{2};p'_{3},p'_{1})(2\pi)^32E_{3}\delta^3(\vec{p}_{3}-\vec{p}'_{2})
$$
\begin{equation}
+i\mathcal{M}_1(p_{1},p_{2};p'_{2},p'_{3})(2\pi)^32E_{3}\delta^3(\vec{p}_{3}-\vec{p}'_{1})\Big) ~+~ (2 \to 3) ~+~(1 \to 3)
\end{equation}
Upon taking the absolute value squared of this amplitude, we get terms that are perfect squares, like for instance,
\begin{equation}
\Big|i\mathcal{M}_1(p_{1},p_{2};p'_{1},p'_{2})\Big|^2\Big[(2\pi)^32E_{3}\delta^3(\vec{p}_{3}-\vec{p}'_{3})\Big]^2
\end{equation}
and cross-terms. 

Let us calculate the contribution from the perfect square above. We work in the center of mass frame of the interacting particles $1,2$. The center of mass energy is denoted by $E_{12}=2E_1$, the relative velocity of the incoming particles by $v_{12}=4|\vec{p}_1|/E_{12}$ and the scattering angle by $\theta$.  Integrating over the final momenta $\vec{p}'_3, \vec{p}'_2$, and $|\vec{p}'_1|$, produces a factor $V^2Tv_{12}/4$, giving a net contribution  
\begin{equation}\label{perfectsquare}
-\frac{T\ln{e^6}}{96V} \ln\left(\frac{E_d}{\lambda}\right) \int_0^\pi d\theta \sin\theta \,\frac{v_{12}}{E_{12}^2} ~\Big|i\mathcal{M}_1(E_{12},\theta)\Big|^2\,\mathcal{B}_1(12;1'2')
\end{equation}
Notice that there are three such terms, for each pair of interacting particles, in accordance with \ref{entmaster}.

There are cross-terms with incompatible $\delta$-functions, for example,
\begin{equation}
\Big [\mathcal{M}_1(p_{1},p_{2};p'_{1},p'_{2})\mathcal{M}^*_1(p_{1},p_{3};p'_{2},p'_{1}) + c.c.\Big]\Big[(2\pi)^32E_{3}\delta^3(\vec{p}_{3}-\vec{p}'_{3})\Big]\Big[(2\pi)^32E_{2}\delta^3(\vec{p}_{2}-\vec{p}'_{3})\Big]
\end{equation}
Upon integration over $\vec{p}'_3$, we get a factor $(2\pi)^3\delta^3(\vec{p}_{3}-\vec{p}_{2})$, which vanishes for generic states in which the incoming three particles have different momenta. Another example that gives a vanishing contribution is the following
\begin{equation}
\Big [\mathcal{M}_1(p_{1},p_{2};p'_{1},p'_{2})\mathcal{M}^*_1(p_{1},p_{2};p'_{3},p'_{1}) + c.c.\Big]\Big[(2\pi)^32E_{3}\delta^3(\vec{p}_{3}-\vec{p}'_{3})\Big]\Big[(2\pi)^32E_{3}\delta^3(\vec{p}_{3}-\vec{p}'_{2})\Big]
\end{equation}
We work in the center of mass frame of the two interacting particles (1 and 2). Integrating over $\vec{p}'_3, \vec{p}'_2$ imposes $\vec{p}'_3=\vec{p}'_2=\vec{p}_3$. This is incompatible with momentum conservation unless $\vec{p}_3=0$. But then generically the energy will not be conserved.

A cross-term with a non-vanishing but suppressed contribution in the large volume, continuum limit is 
\begin{equation}
\Big [\mathcal{M}_1(p_{1},p_{2};p'_{1},p'_{2})\mathcal{M}^*_1(p_{1},p_{3};p'_{1},p'_{3}) + c.c.\Big]\Big[(2\pi)^32E_{3}\delta^3(\vec{p}_{3}-\vec{p}'_{3})\Big]\Big[(2\pi)^32E_{2}\delta^3(\vec{p}_{2}-\vec{p}'_{2})\Big]
\end{equation}
Integration over $\vec{p}'_3, \vec{p}'_2$ imposes $\vec{p}'_3=\vec{p}_3$ and $\vec{p}'_2=\vec{p}_2$. Integration over $\vec{p}'_1$, using one of the overall momentum $\delta$-functions imposes $\vec{p}'_1= \vec{p}_1$ and yields a factor of $VT^2$. The contribution of this term is of order $T^2/(V^2E_1^2E_2E_3)$ times dimensionless amplitudes and kinematical factors. Letting $T \simeq L/c$, where $L$ is the size of the box and $c$ the speed of light, the ratio of this contribution to \ref{perfectsquare} is of order $1/(L^2 E_2 E_3)$, which becomes vanishingly small when the IR cutoff $1/L$ is taken to zero.   

\section{Traces to next to leading order}\label{a4}
\setcounter{equation}{0}
For the next to leading order computation concerning the scattering of two Fock basis electrons, the reduced density matrix takes the explicit form
\begin{gather}
\rho_H=\Tr_{H_S}\ket{\Psi}_{out}\bra{\Psi}_{out}=\nonumber\\
\ket{\alpha}_H\bra{\alpha}_H+\bigg(T_{\alpha\beta}\ket{\beta}_H+\sum_{\omega_\gamma>E}T_{\beta\gamma,\alpha}\ket{\beta\gamma}_H+\sum_{\ket{\gamma_1\gamma_2}\in{\cal{H}}_H}T_{\beta\gamma_1\gamma_2,\alpha}\ket{\beta\gamma_1\gamma_2}_H+\dots\bigg)\bra{\alpha}_H\nonumber\\
+\ket{\alpha}_H\bigg(T^*_{\alpha\beta}\bra{\beta}_H+\sum_{\omega_{\gamma'}>E}T^*_{\beta'\gamma',\alpha}\bra{\beta'\gamma'}_H+\sum_{\ket{\gamma'_1\gamma'_2}\in{\cal{H}}_H}T^*_{\beta'\gamma'_1\gamma'_2,\alpha}\bra{\beta'\gamma'_1\gamma'_2}_H+\dots\bigg) \nonumber\\
+D_{\beta,\beta'}\ket{\beta}_H\bra{\beta'}_H+\sum_{\omega_{\gamma'}>E}D_{\beta,\beta'\gamma'}\ket{\beta}_H\bra{\beta'\gamma'}_H+\sum_{\omega_{\gamma}>E}D_{\beta\gamma,\beta'}\ket{\beta\gamma}_H\bra{\beta'}_H\nonumber\\
+\sum_{\omega_\gamma,\omega_{\gamma'}>E}D_{\beta\gamma,\beta'\gamma'}\ket{\beta\gamma}_H\bra{\beta'\gamma'}_H \nonumber\\
+\sum_{\ket{\gamma'_1\gamma'_2}\in{\cal{H}}_H}T_{\beta\alpha}T^*_{\beta'\gamma'_1\gamma'_2,\alpha}\ket{\beta}_H\bra{\beta'\gamma'_1\gamma'_2}_H+\sum_{\ket{\gamma_1\gamma_2}\in{\cal{H}}_H}T^*_{\beta'\alpha}T_{\beta\gamma_1\gamma_2,\alpha}\ket{\beta\gamma_1\gamma_2}_H\bra{\beta'}_H+\dots\nonumber\\
+\sum_{\substack{\ket{\gamma'_1\gamma'_2}\in{\cal{H}}_H\\\omega_\gamma>E}}T_{\beta\gamma,\alpha}T^*_{\beta'\gamma'_1\gamma'_2,\alpha}\ket{\beta\gamma}_H\bra{\beta'\gamma'_1\gamma'_2}_{H}+\sum_{\substack{\ket{\gamma_1\gamma_2}\in{\cal{H}}_H\\\omega_{\gamma'}>E}}T^*_{\beta'\gamma',\alpha}T_{\beta\gamma_1\gamma_2,\alpha}\ket{\beta\gamma_1\gamma_2}_{H}\bra{\beta'\gamma'}_H+\dots \nonumber\\
+\sum_{\substack{\ket{\gamma_1\gamma_2}\in{\cal{H}}_H\\\ket{\gamma'_1\gamma'_2}\in{\cal{H}}_H}}T^*_{\beta'\gamma'_1\gamma'_2,\alpha}T_{\beta\gamma_1\gamma_2,\alpha}\ket{\beta\gamma_1\gamma_2}_{H}\bra{\beta'\gamma'_1\gamma'_2}_H+\dots
\end{gather}
where
\begin{gather}
D_{\beta,\beta'}=T_{\beta\alpha}T^*_{\beta'\alpha}+\sum_{\omega_\gamma<E}T_{\beta\gamma,\alpha}T^*_{\beta'\gamma,\alpha}+\sum_{\ket{\gamma_1\gamma_2}\in{\cal{H}}_S}T^*_{\beta'\gamma_1\gamma_2,\alpha}T_{\beta\gamma_1\gamma_2,\alpha}+\dots
\end{gather}
\begin{gather}
D_{\beta\gamma,\beta'}=T^*_{\beta'\alpha}T_{\beta\gamma,\alpha}+\sum_{\substack{\omega_{\gamma'}<E}}T^*_{\beta'\gamma',\alpha}T_{\beta\gamma\gamma',\alpha}+\dots
\end{gather}
\begin{gather}
D_{\beta,\beta'\gamma'}=T_{\beta\alpha}T^*_{\beta'\gamma',\alpha}+\sum_{\substack{\omega_{\gamma}<E}}T_{\beta\gamma,\alpha}T^*_{\beta'\gamma\gamma',\alpha}+\dots
\end{gather}
and
\begin{gather}
D_{\beta\gamma,\beta'\gamma'}=T_{\beta\gamma,\alpha}T^*_{\beta'\gamma',\alpha}+\sum_{\substack{\omega_{\gamma_1}<E}}T_{\beta\gamma\gamma_1,\alpha}T^*_{\beta'\gamma'\gamma_1,\alpha}+\dots
\end{gather}
A sum over $\beta,\, \beta'$ is implied. The ellipses in the equations above stand for terms of order greater than $e^{8}$. 

The unitarity relation takes the form
\begin{gather}
T_{\alpha\alpha}+T^*_{\alpha\alpha}+\sum_{\beta}T_{\beta\alpha}T^*_{\beta\alpha}+\sum_{\beta,\gamma}T_{\beta\gamma,\alpha}T^*_{\beta\gamma,\alpha}+\sum_{\substack{\beta,\gamma_1,\gamma_2\\\omega_{\gamma_1}\leq\omega_{\gamma_2}}}T_{\beta\gamma_1\gamma_2,\alpha}T^*_{\beta\gamma_1\gamma_2,\alpha}+\dots=0
\end{gather}
From this relation we conclude that $T_{\alpha\alpha}+T^*_{\alpha\alpha}$ is of order $e^{4}$ and $T_{\alpha\alpha}+T^*_{\alpha\alpha}+\sum_{\beta}T_{\beta\alpha}T^*_{\beta\alpha}$ of order $e^{6}$. 

We write the density matrix in the form $\rho_H=\ket{\alpha}\bra{\alpha}+\epsilon\equiv\rho_0+\varepsilon$, where $\varepsilon$  is a perturbation of order $e^{2}$. So we need to expand the powers of the reduced density matrix to order $\epsilon^4$ in order to calculate the Renyi entropies for integer $m\ge 2$ to order $e^{8}$. Due to the cyclic property of the trace, it suffices to consider the traces of the following structures: $\rho_0\varepsilon$,  $ \varepsilon^2$, $\rho_0\varepsilon^2$, $ \rho_0\varepsilon\rho_0\varepsilon$, $\varepsilon^3$, $\rho_0\varepsilon^3$, $ \rho_0\varepsilon^2\rho_0\varepsilon$, $(\rho_0\varepsilon)^3$, $\varepsilon^4$, $\rho_0\varepsilon^4$, $\rho_0\varepsilon^3\rho_0\varepsilon$, $\rho_0\varepsilon^2\rho_0\varepsilon^2$, $\rho_0\varepsilon^2(\rho_0\varepsilon)^2$ and $(\rho_0\varepsilon)^4$.

For the linear and quadratic terms, we get to order $e^8$
\begin{gather}
\Tr\rho_0\varepsilon=T_{\alpha\alpha}+T^*_{\alpha\alpha}+D_{\alpha,\alpha}\nonumber\\
=T_{\alpha\alpha}+T^{*}_{\alpha\alpha}+T_{\alpha\alpha}T^{*}_{\alpha\alpha}+\sum_{\omega_\gamma<E}T_{\alpha\gamma,\alpha}T^*_{\alpha\gamma,\alpha}+\sum_{\ket{\gamma_1\gamma_2}\in{\cal{H}}_S}T_{\alpha\gamma_1\gamma_2,\alpha}T^*_{\alpha\gamma_1\gamma_2,\alpha}
\label{re}    
\end{gather}
\begin{gather}
\Tr\varepsilon^2
=T_{\alpha\alpha}^2+T^{*2}_{\alpha\alpha}+2T_{\beta\alpha}T^*_{\beta\alpha}+2\sum_{\omega\gamma>E}T_{\beta\gamma,\alpha}T^*_{\beta\gamma,\alpha}+2\sum_{\ket{\gamma_1\gamma_2}\in{\cal{H}}_H}T_{\beta\gamma_1\gamma_2,\alpha}T^*_{\beta\gamma_1\gamma_2,\alpha}
\nonumber\\
+2T_{\beta\alpha} D_{\alpha,\beta}+2T^*_{\beta\alpha} D_{\beta,\alpha}+2\sum_{\omega_\gamma>E}T_{\beta\gamma,\alpha}D_{\alpha,\beta\gamma}+2\sum_{\omega_\gamma>E}T^*_{\beta\gamma,\alpha}D_{\beta\gamma,\alpha}+D_{\beta,\beta'}D_{\beta',\beta}
\nonumber\\
=T_{\alpha\alpha}^2+T^{*2}_{\alpha\alpha}+2T_{\beta\alpha}T^*_{\beta\alpha}+2\sum_{\omega_\gamma>E}T_{\beta\gamma,\alpha}T^*_{\beta\gamma,\alpha}+2\sum_{\ket{\gamma_1\gamma_2}\in{\cal{H}}_H}T_{\beta\gamma_1\gamma_2,\alpha}T^*_{\beta\gamma_1\gamma_2,\alpha}
\nonumber\\
+2\Big(T_{\alpha\alpha}+T^*_{\alpha\alpha}\Big)T_{\beta\alpha}T^*_{\beta\alpha}+ (T_{\beta\alpha}T^*_{\beta\alpha})^2
\label{e2}    
\end{gather}
\begin{gather}
\Tr\rho_0\varepsilon^2
=T_{\alpha\alpha}^2+T_{\alpha\alpha}T^{*}_{\alpha\alpha}+T^{*2}_{\alpha\alpha}
+T_{\beta\alpha}T^*_{\beta\alpha}\nonumber\\
+(T_{\alpha\alpha}+T^{*}_{\alpha\alpha})D_{\alpha,\alpha}+T_{\beta\alpha} D_{\alpha,\beta}+T^*_{\beta\alpha} D_{\beta,\alpha}
+\sum_{\omega_\gamma>E}T_{\beta\gamma,\alpha}T^*_{\beta\gamma,\alpha}\nonumber\\
+\sum_{\omega_\gamma>E}T_{\beta\gamma,\alpha}D_{\alpha,\beta\gamma}+\sum_{\omega_\gamma>E}T^*_{\beta\gamma,\alpha}D_{\beta\gamma,\alpha}+\sum_{\ket{\gamma_1\gamma_2}\in{\cal{H}}_H}T_{\beta\gamma_1\gamma_2,\alpha}T^*_{\beta\gamma_1\gamma_2,\alpha}+D_{\alpha,\beta}D_{\beta,\alpha}
\nonumber\\
=T_{\alpha\alpha}^2+T_{\alpha\alpha}T^{*}_{\alpha\alpha}+T^{*2}_{\alpha\alpha}
+T_{\beta\alpha}T^*_{\beta\alpha}(1+T_{\alpha\alpha}+T^{*}_{\alpha\alpha})
\nonumber\\
+\sum_{\omega_\gamma>E}T_{\beta\gamma,\alpha}T^*_{\beta\gamma,\alpha}
+\sum_{\ket{\gamma_1\gamma_2}\in{\cal{H}}_H}T_{\beta\gamma_1\gamma_2,\alpha}T^*_{\beta\gamma_1\gamma_2,\alpha}+T_{\beta\alpha}T^*_{\beta\alpha}T_{\alpha\alpha}T^*_{\alpha\alpha}
\label{rhoe2}    
\end{gather}
and finally
\begin{gather}
\Tr\rho_0\varepsilon\rho_0\varepsilon=T_{\alpha\alpha}^2+2T_{\alpha\alpha}T^{*}_{\alpha\alpha}+T^{*2}_{\alpha\alpha}+2(T_{\alpha\alpha}+T^{*}_{\alpha\alpha})D_{\alpha,\alpha}+D_{\alpha,\alpha}^2\nonumber\\
=(T_{\alpha\alpha}+T^{*}_{\alpha\alpha})(T_{\alpha\alpha}+T^{*}_{\alpha\alpha}+2T_{\alpha\alpha}T^{*}_{\alpha\alpha})+T^2_{\alpha\alpha}T^{*2}_{\alpha\alpha}
\label{erer}    
\end{gather}

To cubic order we have
\begin{gather}
\Tr\varepsilon^3=T_{\alpha\alpha}^3+T_{\alpha\alpha}^{*3}+3(T_{\alpha\alpha}+T^{*}_{\alpha\alpha}+T_{\beta'\alpha}T^*_{\beta'\alpha}+T_{\alpha\alpha}^2+T_{\alpha\alpha}T^{*}_{\alpha\alpha}+T^{*2}_{\alpha\alpha})T_{\beta\alpha}T^*_{\beta\alpha}\nonumber\\
+3\sum_{\omega_\gamma>E}(T_{\alpha\alpha}+T^{*}_{\alpha\alpha})T_{\beta\gamma,\alpha}T^*_{\beta\gamma,\alpha}
=T_{\alpha\alpha}^3+T_{\alpha\alpha}^{*3}+3T_{\beta\alpha}T^*_{\beta\alpha}(T_{\alpha\alpha}^2+T_{\alpha\alpha}T^{*}_{\alpha\alpha}+T^{*2}_{\alpha\alpha})
=-2(T_{\alpha\alpha}+T^{*}_{\alpha\alpha})^3=0
\label{e3}    
\end{gather}
\begin{gather}
\Tr\rho_0\varepsilon^3=T_{\alpha\alpha}^3+T_{\alpha\alpha}^{*3}+T_{\alpha\alpha}^2T^{*}_{\alpha\alpha}+T_{\alpha\alpha}T^{*2}_{\alpha\alpha}+T_{\alpha\alpha}^3T^{*}_{\alpha\alpha}+T_{\alpha\alpha}^2T^{*2}_{\alpha\alpha}+T_{\alpha\alpha} T_{\alpha\alpha}^{*3}\nonumber\\
+T_{\beta\alpha}T^*_{\beta\alpha}(2T_{\alpha\alpha}+2T^{*}_{\alpha\alpha}+4T_{\alpha\alpha}T^{*}_{\alpha\alpha}+2T_{\alpha\alpha}^2+2T^{*2}_{\alpha\alpha}+T_{\beta'\alpha}T^*_{\beta'\alpha})
+2\sum_{\omega_\gamma>E}(T_{\alpha\alpha}+T^{*}_{\alpha\alpha})T_{\beta\gamma,\alpha}T^*_{\beta\gamma,\alpha}\nonumber\\
=(T_{\alpha\alpha}+T^{*}_{\alpha\alpha}+T_{\alpha\alpha}T^{*}_{\alpha\alpha})(T_{\alpha\alpha}^2+T^{*2}_{\alpha\alpha})+T_{\alpha\alpha}^2T^{*2}_{\alpha\alpha}+(T_{\alpha\alpha}+T^{*}_{\alpha\alpha})T_{\beta\alpha}T^*_{\beta\alpha}
\label{re3}    
\end{gather}
\begin{gather}
\Tr\rho_0\varepsilon^2\rho_0\varepsilon=T_{\alpha\alpha}^3+T_{\alpha\alpha}^{*3}+2T_{\alpha\alpha}^2T^{*}_{\alpha\alpha}+2T_{\alpha\alpha}T^{*2}_{\alpha\alpha}+3T_{\alpha\alpha}^2T^{*2}_{\alpha\alpha}+2T_{\alpha\alpha}^3T^{*}_{\alpha\alpha}+2T_{\alpha\alpha} T_{\alpha\alpha}^{*3}\nonumber\\
+T_{\beta\alpha}T^*_{\beta\alpha}(T_{\alpha\alpha}+T^{*}_{\alpha\alpha}+T_{\alpha\alpha}^2+T^{*2}_{\alpha\alpha}+3T_{\alpha\alpha}T^{*}_{\alpha\alpha})+\sum_{\omega_\gamma>E}(T_{\alpha\alpha}+T^{*}_{\alpha\alpha})T_{\beta\gamma,\alpha}T^*_{\beta\gamma,\alpha}\nonumber\\
=(T_{\alpha\alpha}+T^{*}_{\alpha\alpha}+T_{\alpha\alpha}T^{*}_{\alpha\alpha})(T_{\alpha\alpha}^2+T^{*2}_{\alpha\alpha}+T_{\alpha\alpha}T^{*}_{\alpha\alpha}+T_{\beta\alpha}T^*_{\beta\alpha})
\label{re2re}    
\end{gather}
\begin{gather}
\Tr(\rho_0\varepsilon)^3=T_{\alpha\alpha}^3+T_{\alpha,\alpha}^{*3}+3T_{\alpha\alpha}T^{*}_{\alpha\alpha}(T_{\alpha\alpha}+T^{*}_{\alpha\alpha}+T_{\alpha\alpha}^2+T^{*2}_{\alpha\alpha}+2T_{\alpha\alpha}T^{*}_{\alpha\alpha})
\nonumber\\
=(T_{\alpha\alpha}+T^{*}_{\alpha\alpha})^3+3T_{\alpha\alpha}T^{*}_{\alpha\alpha}(T_{\alpha\alpha}+T^{*}_{\alpha\alpha})^2=0
\label{re3}    
\end{gather}
The latter trace does not contribute to order $e^{8}$.

The quartic terms yield 
\begin{gather}
\Tr\varepsilon^4
=T_{\alpha\alpha}^4+T_{\alpha\alpha}^{*4}+2T_{\beta\alpha}T^*_{\beta\alpha}(T_{\alpha\alpha}^2+T^{*2}_{\alpha\alpha}+T_{\beta'\alpha}T^*_{\beta'\alpha})
\label{e4}    
\end{gather}
\begin{gather}
\Tr\rho_0\varepsilon^4
=T_{\alpha\alpha}^2T^{*2}_{\alpha\alpha}+T_{\beta\alpha}T^*_{\beta\alpha}(T_{\alpha\alpha}^2+T^{*2}_{\alpha\alpha}+T_{\beta'\alpha}T^*_{\beta'\alpha})
\label{re4}    
\end{gather}
\begin{gather}
\Tr\rho_0\varepsilon^2\rho_0\varepsilon^2
=T_{\alpha\alpha}^2T^{*2}_{\alpha\alpha}+T_{\beta\alpha}T^*_{\beta\alpha}(T_{\alpha\alpha}^2+T^{*2}_{\alpha\alpha}+T_{\beta'\alpha}T^*_{\beta'\alpha})
\label{re2re2}    
\end{gather}
and 
\begin{equation}
\Tr\rho_0\varepsilon^3\rho_0\varepsilon=\Tr\rho_0\varepsilon^2(\rho_0\varepsilon)^2=\Tr(\rho_0\varepsilon)^4=0
\end{equation}

We proceed now to compute $\Tr(\rho_H)^m$ for integer $m\ge 2$.  For $m=2$, we obtain to order $e^8$
\begin{gather}
\Tr(\rho_H)^2=1+2\Tr\rho_0\varepsilon+\Tr\varepsilon^2\nonumber\\\nonumber\\
=1+2(T_{\alpha\alpha}+T^{*}_{\alpha\alpha}+T_{\beta\alpha}T^*_{\beta\alpha})
+2\sum_{\omega_\gamma>E}T_{\beta\gamma,\alpha}T^*_{\beta\gamma,\alpha}+2\sum_{\ket{\gamma_1\gamma_2}\in {\cal{H}}_H}T_{\beta\gamma_1\gamma_2,\alpha}T^*_{\beta\gamma_1\gamma_2,\alpha}
\nonumber\\
=1-2\sum_{\omega_\gamma<E}T_{\beta\gamma,\alpha}T^*_{\beta\gamma,\alpha}-2\sum_{\substack{\ket{\gamma_1\gamma_2}\ni{\cal{H}}_H}}T_{\beta\gamma_1\gamma_2,\alpha}T^*_{\beta\gamma_1\gamma_2,\alpha}=1-2\Delta
 \end{gather}
where 
 \begin{equation}
 \Delta=\sum_{\omega_\gamma<E}T_{\beta\gamma,\alpha}T^*_{\beta\gamma,\alpha}-\sum_{\substack{\ket{\gamma_1\gamma_2}\ni{\cal{H}}_H}}T_{\beta\gamma_1\gamma_2,\alpha}T^*_{\beta\gamma_1\gamma_2,\alpha}
 \end{equation}
To arrive to this result we applied the unitarity relation. It is important to emphasise that the second sum in the expression for $\Delta$ includes cases for which one photon is soft while the other is hard.

For the cubic and the quartic traces, we get
\begin{gather}
\Tr(\rho_H)^3=1+3\Tr\rho_0\varepsilon+3\Tr\rho_0\varepsilon^2+\Tr\varepsilon^3
\nonumber\\
=1+3(T_{\alpha\alpha}+T^{*}_{\alpha\alpha}+T_{\beta\alpha}T^*_{\beta\alpha})+3(T_{\alpha\alpha}+T^{*}_{\alpha\alpha})^2+3T_{\beta\alpha}T^*_{\beta\alpha}(T_{\alpha\alpha}+T^{*}_{\alpha\alpha})\nonumber\\
+3\sum_{\omega_\gamma>E}T_{\beta\gamma,\alpha}T^*_{\beta\gamma,\alpha}+3\sum_{\ket{\gamma_1\gamma_2}\in{\cal{H}}_H}T_{\beta\gamma_1\gamma_2,\alpha}T^*_{\beta\gamma_1\gamma_2,\alpha}
+T_{\alpha\alpha}^3+T_{\alpha\alpha}^{*3}+T_{\beta\alpha}T^*_{\beta\alpha}(A_{\alpha\alpha}^2+T^{*2}_{\alpha\alpha}+T_{\alpha\alpha}T^{*}_{\alpha\alpha})
\nonumber\\
=1-3\sum_{\omega_\gamma<E}T_{\beta\gamma,\alpha}T^*_{\beta\gamma,\alpha}-3\sum_{\substack{\ket{\gamma_1\gamma_2}\ni{\cal{H}}_H}}T_{\beta\gamma_1\gamma_2,\alpha}T^*_{\beta\gamma_1\gamma_2,\alpha}=1-3\Delta
\end{gather}
\begin{gather}
\Tr(\rho_H)^4=1+4\Tr\rho_0\varepsilon+4\Tr\rho_0\varepsilon^2+2\Tr\rho_0\varepsilon\rho_0\varepsilon+4\Tr\rho_0\varepsilon^3+\Tr\varepsilon^4
\nonumber\\
=1+4\bigg(T_{\alpha\alpha}+T^{*}_{\alpha\alpha}+T_{\beta\alpha}T^*_{\beta\alpha}+\sum_{\omega_\gamma>E}T_{\beta\gamma,\alpha}T^*_{\beta\gamma,\alpha}+\sum_{\ket{\gamma_1\gamma_2}\in{\cal{H}}_H}T_{\beta\gamma_1\gamma_2,\alpha}T^*_{\beta\gamma_1\gamma_2,\alpha}\bigg)
\nonumber\\
+2(T_{\alpha\alpha}+T^{*}_{\alpha\alpha})(3T_{\alpha\alpha}+3T^{*}_{\alpha\alpha}+2T_{\alpha\alpha}^2+2T^{*2}_{\alpha\alpha}+2T_{\alpha\alpha}T^{*}_{\alpha\alpha}+4T_{\beta\alpha}T^{*}_{\beta\alpha})
\nonumber\\
+2T_{\alpha\alpha}T^{*}_{\alpha\alpha}(2T_{\alpha\alpha}^2+2T^{*2}_{\alpha\alpha}+3T_{\alpha\alpha}T^{*}_{\alpha\alpha})+T_{\alpha\alpha}^4+T_{\alpha\alpha}^{*4}+2T_{\beta\alpha}T^*_{\beta\alpha}\big(T^2_{\alpha\alpha}+T^{2*}_{\alpha\alpha}+T_{\beta'\alpha}T^*_{\beta'\alpha}\big)
\nonumber\\
=1-4\sum_{\omega_\gamma<E}T_{\beta\gamma,\alpha}T^*_{\beta\gamma,\alpha}-4\sum_{\substack{\ket{\gamma_1\gamma_2}\ni{\cal{H}}_H}}T_{\beta\gamma_1\gamma_2,\alpha}T^*_{\beta\gamma_1\gamma_2,\alpha} = 1-4\Delta
\end{gather}

For $m>4$, we use the following expansion of the trace of $\rho_H$, valid to order $e^{8}$, 
\begin{gather}
\Tr(\rho_H)^m=1+m\Tr\rho_0\varepsilon+m\Tr\rho_0\varepsilon^2+\frac{m(m-3)}{2}\Tr\rho_0\varepsilon\rho_0\varepsilon+m\Tr\rho_0\varepsilon^3
\nonumber\\
+m(m-4)\Tr\rho_0\varepsilon^2\rho_0\varepsilon+m\Tr\rho_0\varepsilon^4+\frac{m(m-5)}{2}\Tr\rho_0\varepsilon^2\rho_0\varepsilon^2
 \end{gather}
to obtain 
\begin{gather}
\Tr(\rho_H)^m
=1+m\bigg(T_{\alpha\alpha}+T^{*}_{\alpha\alpha}+T_{\beta\alpha}T^*_{\beta\alpha}+\sum_{\omega_\gamma>E}T_{\beta\gamma,\alpha}T^*_{\beta\gamma,\alpha}+\sum_{\ket{\gamma_1\gamma_2}\in{\cal{H}}_H}T_{\beta\gamma_1\gamma_2,\alpha}T^*_{\beta\gamma_1\gamma_2,\alpha}\bigg)
\nonumber\\
=1-m\sum_{\omega_\gamma<E}T_{\beta\gamma,\alpha}T^*_{\beta\gamma,\alpha}-m\sum_{\substack{\ket{\gamma_1\gamma_2}\ni{\cal{H}}_H}}T_{\beta\gamma_1\gamma_2,\alpha}T^*_{\beta\gamma_1\gamma_2,\alpha}=1-m\Delta
\end{gather}
 
\section{Next to leading order corrections to the eigenvalues of $\rho_H$}\label{a5}
In this Appendix we describe how to obtain next to leading order corrections to the order $e^6$ eigenvalues of $G$ using second order perturbation theory. To this extend we write
\begin{equation}
G=G^{(0)}+G^{(1)}
\end{equation}
where
\begin{equation}
G^{(0)}=\bigg(\sum_{\omega_\gamma<E}T_{\beta\gamma,\alpha}T^*_{\beta'\gamma,\alpha}+\sum_{\ket{\gamma_1\gamma_2}\in {\cal{H}}_S}T_{\beta\gamma_1\gamma_2,\alpha}T^*_{\beta'\gamma_1\gamma_2,\alpha}\bigg)\ket{\beta}_H\bra{\beta'}_H ~+~\sum_{\substack{\omega_{\gamma_1}<E}}T_{\beta\gamma\gamma_1,\alpha}T^*_{\beta'\gamma'\gamma_1,\alpha}\ket{\beta\gamma}_H\bra{\beta'\gamma'}_H 
\end{equation}
and treat
\begin{equation}
G^{(1)}=\sum_{\substack{\omega_{\gamma}<E}}T_{\beta\gamma,\alpha}T^*_{\beta'\gamma\gamma',\alpha}\ket{\beta}_H\bra{\beta'\gamma'}_H + h.c.
\end{equation}
as a perturbation to find the eigenvalues of $G$. Since $G^{(0)}$ assumes a block diagonal form, it does not mix states with no photons, of the form $\ket{\beta}$, with the single-photon states $\ket{\beta\gamma}$. Because of energy conservation, we may restrict to the subspace spanned by states with energies $E_\alpha>E_\beta>E_\alpha-E$ and $E_\alpha>E_{\beta\gamma}>E_\alpha-E$. These states are orthogonal to $\ket{\Phi}$ (and so are annihilated by $\ket{\Phi}\bra{\Phi}$). In particular the eigenstates of $G$ with non-vanishing eigenvalues necessarily lie in this orthogonal to $\ket{\Phi}$ subspace, and so are simultaneously eigenstates of $\rho_H$ (with the same eigenvalue). 

The eigenstates of $G^{(0)}$ with eigenvalues of order $e^6$ comprise linear combinations of two-electron states with no photons (and energies satisfying $E_\alpha>E_\beta>E_\alpha-E$) only. We denote these order $e^6$ eigenstates by $\ket{\tilde{\beta}}$ and the corresponding eigenvalues by $\lambda_{\tilde \beta}^{(0)}$. The sum of these eigenvalues is given by $\sum \lambda_{\tilde \beta}^{(0)}=\Delta_6$, where $\Delta_6$ stands for the order $e^6$ part of $\Delta$. 

There are also linear combinations of $\ket{\beta}$ states that are eigenstates of $G^{(0)}$ with eigenvalues of order $e^8$ (or higher). For example, up to order $e^6$, momentum conservation requires that two-electron $\ket{\beta}$ states with vanishing net momentum are annihilated by $G^{(0)}$. Presumably, certain linear combinations of such zero-momentum two-electron states can be eigenstates of $G^{(0)}$ with eigenvalues of order $e^8$. Certain linear combinations of $\ket{\beta}$ states involving three electrons and a positron can also be eigenstates of $G^{(0)}$ with eigenvalues of order $e^8$. Notice however that the perturbation $G^{(1)}$ does not mix these eigenstates with order $e^6$ eigenstates, or the order $e^6$ eigenstates among themselves. To find the $e^8$ corrections to the leading eigenstates of $G^{(0)}$, we must apply second order perturbation theory.

The other eigenstates of $G^{(0)}$, with eigenvalues of order $e^8$, comprise linear combinations of single photon states (and energies $E_\alpha>E_{\beta\gamma}>E_\alpha-E$) only. We shall denote these eigenstates by $\ket{\widetilde{\beta\gamma}}$ and the corresponding eigenvalues by $\lambda_{\widetilde {\beta\gamma}}^{(0)}$. The perturbation $G^{(1)}$ mixes these eigenstates with the order $e^6$ eigenstates of $G^{(0)}$. 

Assuming that the order $e^6$ eigenstates of $G^{(0)}$ have been found to be non-degenerate, we can apply second order perturbation theory to determine the order $e^8$ corrections to the eigenvalues
\begin{equation}
\delta \lambda_{\tilde \beta}^{(0)}=\sum_{\widetilde{\beta\gamma}} {\left|\bra{\tilde\beta}G^{(1)}\ket{\widetilde{\beta\gamma}}\right|^2}/\left(\lambda_{\tilde{\beta}}^0-\lambda_{\widetilde{\beta\gamma}}^0\right)=\sum_{\widetilde{\beta\gamma}}~(\lambda_{\tilde{\beta}}^0)^{-1}~ \bigg|\sum_{\substack{\omega_{\gamma}<E}}T_{\beta\gamma,\alpha}T^*_{\beta'\gamma\gamma',\alpha}\langle{\tilde \beta}|{\beta}\rangle\bra{\beta'\gamma'}{\widetilde{\beta\gamma}}\rangle\bigg|^2 ~+~{\cal{O}}(e^{10})\label{shift1}
\end{equation}



\end{document}